\documentclass[fleqn,usenatbib]{mnras}

\usepackage{newtxtext,newtxmath}

\usepackage[T1]{fontenc}

\DeclareRobustCommand{\VAN}[3]{#2}
\let\VANthebibliography\thebibliography
\def\thebibliography{\DeclareRobustCommand{\VAN}[3]{##3}\VANthebibliography}

\usepackage{graphicx}	
\usepackage{amsmath}	

\title[Stellar-induced differentiation]{Post-main sequence thermal evolution of planetesimals}

\author[Yuqi Li et al.]{
Yuqi Li,$^{1}$\thanks{E-mail: yl817@cam.ac.uk}
Amy Bonsor,$^{1}$
and Oliver Shorttle$^{1,2}$
\\
$^{1}$Institute of Astronomy, University of Cambridge, Madingley Road, Cambridge, CB3 0HA, UK\\
$^{2}$Department of Earth Sciences, University of Cambridge, Downing Street, Cambridge, CB2 3EQ, UK \\
}

\date{Accepted XXX. Received YYY; in original form ZZZ}

\pubyear{2023}

\begin{document}
\setlength{\parskip}{0pt}
\label{firstpage}
\pagerange{\pageref{firstpage}--\pageref{lastpage}}
\maketitle
\begin{abstract}
White dwarfs that have accreted planetary materials provide a powerful tool to probe the interiors and formation of exoplanets. In particular, the high Fe/Si ratio of some white dwarf pollutants suggests that they are fragments of bodies that were heated enough to undergo large-scale melting and iron core formation. In the solar system, this phenomenon is associated with bodies that formed early and so had short-lived radionuclides to power their melting, and/or grew large. However, if the planetary bodies accreted by white dwarfs formed during the (pre)-main sequence lifetime of the host star, they will have potentially been exposed to a second era of heating during the star’s giant branches. This work aims to quantify the effect of stellar irradiation during the giant branches on planetary bodies by coupling stellar evolution to thermal and orbital evolution of planetesimals. We find that large-scale melting, sufficient to form an iron core, can be induced by stellar irradiation, but only in close-in small bodies: planetesimals with radii $\lesssim 30$\,km originally within $\sim2$\,AU orbiting a 1--3\,$M_{\odot}$ host star with solar metallicity. Most of the observed white dwarf pollutants are too massive to be explained by the accretion of these small planetesimals that are melted during the giant branches. Therefore, we conclude that those white dwarfs that have accreted large masses of materials with enhanced or reduced Fe/Si remain an indicator of planetesimal's differentiation shortly after formation, potentially linked to radiogenic heating.
\end{abstract}




\section{Introduction}

\begin{figure*}
\centering
\includegraphics[width=\textwidth]{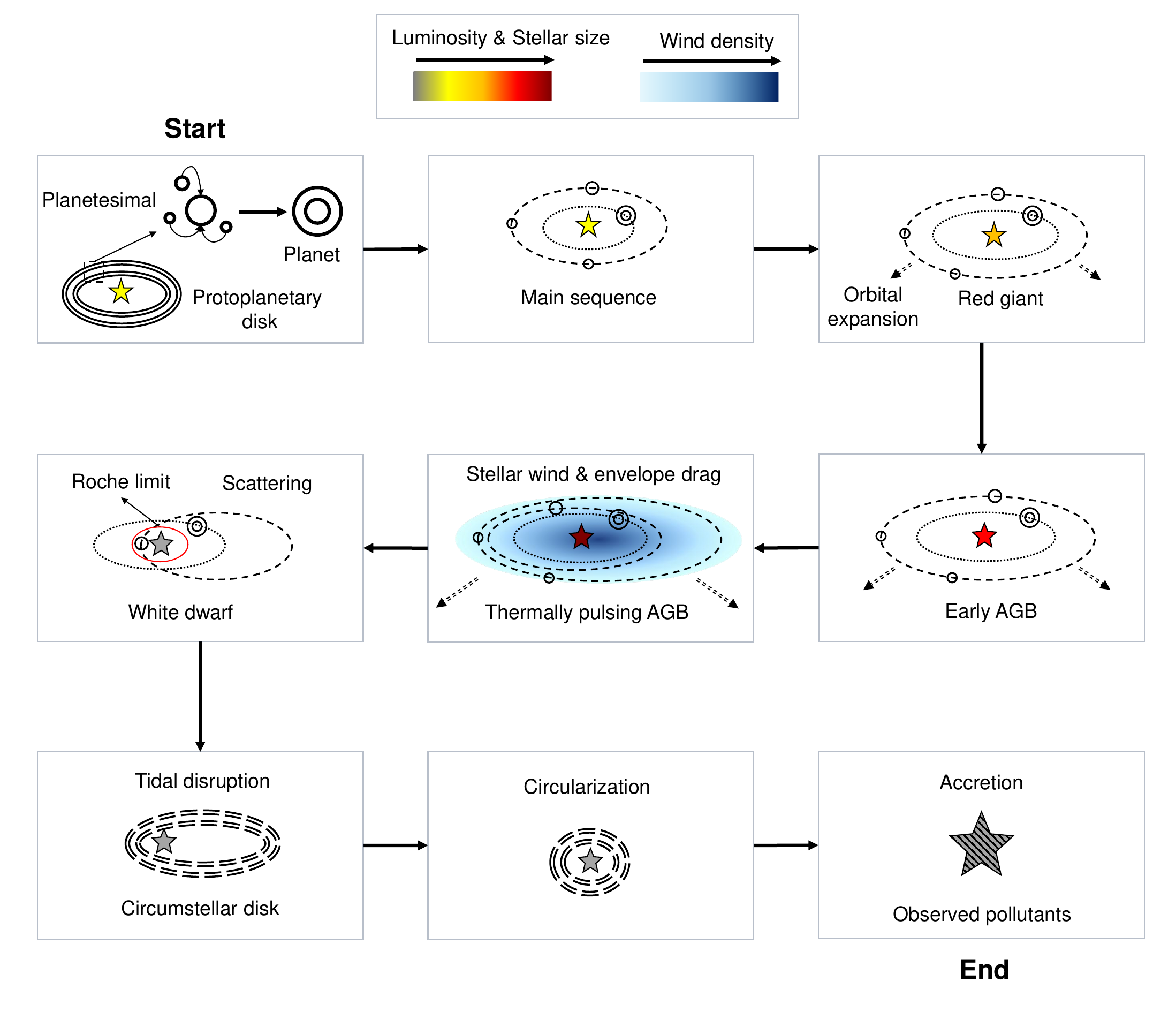}
\caption{Thermal history of white dwarf pollutants: formation from protoplanetary disks, main sequence and post-main sequence life of the star, scattering into the Roche limit, circumstellar disks formation, accretion. The color bars show the general luminosity and radius variations of the host star, and the thermally pulsing asymptotic giant branch stellar wind density.}
\label{thermal history}
\end{figure*}

In an era of exoplanet detection, there is a growing interest in the interior dynamics and volatile content of exoplanets, two properties that are crucial for their habitability. However, our understanding of exoplanet interiors are limited by detection techniques which only reveal their bulk properties, and poorly constrained interior modeling \citep{2015arXiv150203605D,2018arXiv181004615W,2022arXiv220409558W}. Fortunately, white dwarfs (WDs) that have accreted debris of tidally disrupted planetesimals (planetary building blocks) provide a potentially powerful tool to probe the interiors and formation processes of exoplanets. 

White dwarfs, remnants of degenerate stellar cores, should originally preserve atmospheres predominantly composed of hydrogen/helium due to the rapid gravitational settling of heavier elements after radiative levitation becomes negligible (effective temperature $\lesssim20000$\,K). Therefore, the metallic absorption features in the spectra of $\sim$ 20\%--50\% of the WDs possibly originate from recent/ongoing accretions \citep{2003ApJ...596..477Z,2010ApJ...722..725Z,Koester_2014,2019MNRAS.487..133W}. 

The observed WD pollutants reveal diverse compositions for planetary bodies \citep{article}. A large fraction of pollutants resemble the bulk Earth, consistent with tidal disruption and accretion of thermally processed rocky planetesimals \citep{2018MNRAS.479.3814H,2019Sci...366..356D,2021MNRAS.504.2853H,trierweiler2023chondritic}. Exceptions with excess oxygen compared to that hosted in metal oxides indicate the likely accretion of water \citep{2013Sci...342..218F,2020MNRAS.499..171H}. High Ca/Na, Ca/Mn relative to stellar abundances, corresponding to depletion of moderately volatiles is typically assumed to originate from the formation processes of the planetary system, for instance, incomplete condensation and devolatilisation during secondary melting \citep{2003ApJ...591.1220L,2008RSPTA.366.4205O,2014PNAS..11117029P,SIEBERT2018130,2021MNRAS.504.2853H}. A significant dispersion in Fe/Si is usually interpreted as a natural consequence of asynchronously accreted core/mantle-rich fragments from bodies differentiated due to decay of short-lived radioactive elements (e.g., ${}^{26}$Al) \citep{2014AREPS..42...45J,2020MNRAS.492.2683B,2022MNRAS.510.3512B,2022MNRAS.515..395C,2023MNRAS.519.2646B}.

In the solar system, differentiation of planetesimals: large-scale melting and formation of global magma ocean, followed by gravitational segregation of iron from silicates and the formation of an iron core, is thought to be powered by the decay of short-lived radionuclides (e.g., ${}^{26}\rm Al$) and/or violent impacts \citep{2000P&SS...48..887K,2004E&PSL.223..241C}. This magma ocean phase, accompanied with further (moderately) volatile depletion distinct from that of incomplete condensation \citep{2008arXiv0801.1099S,2020ApJ...897...82V}, shapes the interiors of terrestrial planets. 

However, there is no evidence that these early-stage thermal processes inside the solar system planetary bodies are common in the exoplanetary systems. For materials accreted by white dwarfs, late-stage thermal processes induced by stellar evolution, for instance, heating due to stellar irradiation and tidal dissipation, may also result in (moderately) volatile depletion \citep{2010AJ....140.1129J,2016ApJ...832..160M,2017ApJ...842...67M,2017ApJ...849....8M}, large-scale melting and iron core formation, thereby mimicking their early-stage counterparts taking place around planetary formation. To probe the formation stages of a planetary system via WD pollutants, it is thus essential to distinguish between the early-stage and late-stage thermal processes.

This paper models the thermal evolution of planetesimals induced by stellar irradiation and quantifies in what locations and under what conditions planetesimals are sufficiently heated to undergo differentiation (large-scale melting and formation of an iron core) during the giant branches of the host star. As a result, we comment on the possibility that a high level of Fe/Si in the atmosphere of a white dwarf originates from planetary bodies differentiated under stellar irradiation instead of radiogenic heating/impacts. 

\subsection{Journey of planetesimals from formation to accretion onto the white dwarf}

The thermal and orbital evolution of planetesimals accreted by the white dwarf is coupled to stellar evolution (Figure \ref{thermal history}). After the formation of the planetary system, remaining planetesimals experience a long period of steady heating from the main-sequence host star, with its interior temperature approaching a constant value. Afterwards, the interior temperature of planetesimals rises significantly when the host star climbs towards the tip of the red giant branch (RGB, shell hydrogen fusion). Low-mass stars go through helium flash (runaway nuclear fusion) in their degenerate core, while intermediate-mass stars can ignite core helium burning quietly, with the former reaching much higher tip RGB luminosity and radius. Then, stellar luminosity drops, until rising up again near the end of core helium burning (CHB) phase, before entering asymptotic giant branch (AGB, shell helium fusion), when the (average) luminosity keeps rising. Towards the end of AGB, these stars undergo thermal pulsations (TPAGB, cyclical thin shell fusion), accompanied with a rapid increase in average luminosity \citep{1994sipp.book.....H,2015ApJS..219...40C}. At this time, planetesimals experience a short period of intense heating, predominantly affecting a depth equal to the corresponding diffusion length scale. Close-in bodies also experience stronger drag force and/or tidal decay when interacting with the expanding (circumstellar) stellar envelope, spiralling inwards, potentially being disrupted and engulfed by the host star \citep{Maloney_2010,Mustill_2012,2014ApJ...794....3V,Jia_2018}. 

When the host star enters its WD phase after losing a significant fraction of its mass, gravitational perturbation from the planetary system becomes stronger. Planetesimals may be scattered onto highly-eccentric orbits, which in turn decay under tidal interactions with the orbital energy dissipating as heat inside these planetesimals. Finally, once scattered into the Roche limit, planetesimals become tidally disrupted, forming a circumstellar disk, circularized and acrreted onto the WD via a combination of the Poynting-Robertson (PR) effect, gas drag, and the Yarkovsky effect \citep{2011ApJ...732L...3R,2011MNRAS.416L..55R,2011ApJ...741...36B,2012MNRAS.423..505M,https://doi.org/10.48550/arxiv.1505.06204,Malamud_2020,2020MNRAS.493.4692V,2022MNRAS.510.3379V}.

\subsection{Paper layout}

This paper is organized as follows. In Section \ref{method} we summarize the methodology, a synthesis of stellar (\ref{stellar evolution}), planetesimal's orbital (\ref{orbital evolution}) and thermal evolution (\ref{thermal evolution}) codes to obtain the temperature profile evolution of planetesimals, from which we quantify the degree of differentiation (\ref{differentiation fraction}) in a given system. In Section \ref{results}, we present theoretical stellar evolutionary tracks (\ref{stellar evolutionary tracks}), and the resultant temperature profile evolution of sample planetesimals (\ref{temperature profile evolution}). We also present the simulation results of parameter space study in Section \ref{thermal processing inside planetesimals}: the maximum central temperature (\ref{maximum central temperature}) and the degree of differentiation (\ref{volume fraction of melting}) throughout the thermal history of individual planetesimals, with various sizes and orbits. Then, we quantify the mass fraction of planetesimals in a given population that are differentiated as a consequence of stellar evolution until asymptotic giant branch in Section \ref{diff fraction}. In Section \ref{discussion}, we discuss the limitations, their possible alterations to the results (\ref{limitations}), and the implications (\ref{implications}) for observations of white dwarf pollutants. In Section \ref{conclusion}, we summarize this study.

\section{Method}
\label{method}

\begin{figure*}
\includegraphics[width=\textwidth]{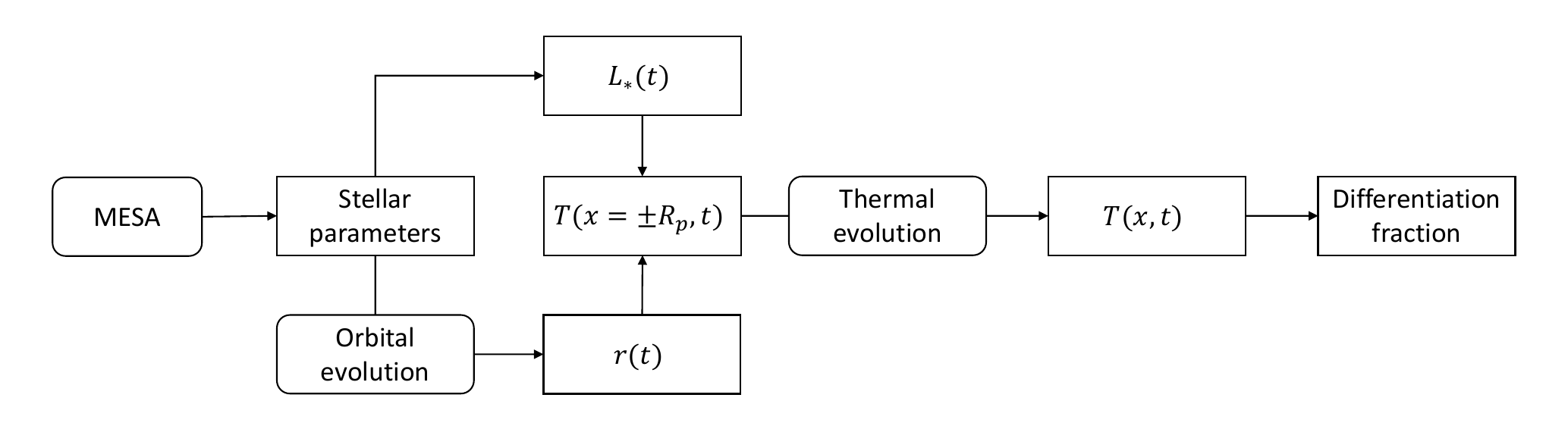}
\caption{Key steps of methodology: a synthesis of stellar and planetesimal's orbital evolution giving the stellar luminosity ($L_*(t)$) and planetesimal's position ($r(t)$), which together provide the stellar irradiation, whose strength is quantified as the planetesimal's surface temperature ($T(x=\pm R_p, t)$). The thermal evolution of the planetesimal is then simulated to predict its internal temperature evolution ($T(x,t)$), which reveals the fraction of the body that undergoes large-scale melting and differentiation. The roundrects and rectangles representing numerical simulations and outputs, respectively. $x$ is the radial coordinate inside the planetesimal, $t$ is the time coordinate.}
\label{method plot}
\end{figure*}

Figure \ref{method plot} summarises our method. The thermal evolution of planetesimals crucially depends on their sizes ($R_p$) and the stellar irradiation. The strength of this irradiation is quantified by the equilibrium temperature at the surface of the planetesimal, a function of stellar luminosity and distance of the body from the host star. First, we model the evolutionary track of the host star utilizing MIST (Section \ref{stellar evolution}), from which we deduce the orbital evolution ($r(t)$) and survival of a planetesimal for a given initial position (\ref{orbital evolution}). Second, based on $r(t)$, and the stellar luminosity variation ($L_*(t)$) from the stellar evolutionary track, we compute the equilibrium surface temperature of the planetesimal ($T(x=\pm R_p,t)$), serving as the boundary condition of the thermal evolution code (\ref{surface equilibrium temperature}). Third, we solve the time-depend temperature profile ($T(x,t)$) inside the planetesimal numerically (\ref{thermal evolution}). Finally, by choosing a critical point $T_{crit}$ such that any region of a planetesimal heated above which is assumed to melt, we quantify the degree of stellar-induced differentiation in planetesimal size ($R_p$)--initial position ($a_0$) space (\ref{thermal processing inside planetesimals}).

\subsection{Stellar evolution}
\label{stellar evolution}

Evolution of the host star plays an important role in the thermal processing and survival of close-in rocky bodies. Stellar luminosity determines the irradiation on planetesimals, whilst stellar mass loss triggers orbital expansion of planetesimals, potentially weakening the irradiation. Additionally, the radius of the star is the key to the survival of planetesimals, as close-in bodies may be engulfed by the stellar envelope.

The set of evolutionary tracks MIST (MESA isochrones and stellar tracks, \citealp{2011ApJS..192....3P,2013ApJS..208....4P,2015ApJS..220...15P,2016ApJS..222....8D,2016ApJ...823..102C}) are used to follow the stellar parameters with a range of initial conditions: initial mass of 1, 2 and 3$\,M_{\odot}$ with an initial metallicity of 0.014. MIST interpolates a grid of single stellar evolutionary tracks from the one dimensional MESA (Modules for Experiments in Stellar Astrophysics) code \citep{2023ApJS..265...15J}, without the need to specify overshooting parameters and fine-tuning of resolutions in order to resolve thermal pulsations of the thermally pulsating AGB (TPAGB) star. MESA solves stellar evolutionary equations utilizing implicit Newton-Raphson method, additionally accounting for time-dependent convection in non-steady state. MESA includes a variety of equation of states, e.g., Skye for fully ionized matter \citep{2021ApJ...913...72J}, FreeEOS for partially ionized matter \citep{2012ascl.soft11002I}. In terms of opacity, MESA accounts for molecular opacity at low temperature region (e.g., close to the surface of TPAGB star), Compton opacity at extremely high temperature and conductive opacity in degenerate region. As a result, MESA is able to model complex post-main sequence evolution of stars, including thermal pulsations driven by nuclear burning in a thin shell.

\subsection{Orbital evolution}
\label{orbital evolution}

The orbital evolution of a planetesimal, together with the stellar luminosity, determines the strength of irradiation on the planetesimal. As the star loses mass, planetary orbits expand. This orbital expansion conserves the specific angular momentum of the planetesimal and reduces the stellar irradiation received (\ref{mass loss}). With the expansion of stellar radius, close-in planetesimals approach and/or enter the stellar envelope. Survival and orbital evolution of these planetesimals depend on their interactions with the environment, via friction, ablation and disruption (\ref{envelope}). We neglect orbital tidal decay (Appendix \ref{tide}, \citealp{10.1093/mnras/stz2339,Mustill_2012}) since it is usually negligible for planetesimal-sized bodies.
 
\subsubsection{Stellar mass loss}
\label{mass loss}

The gravitational force on a close-in planetesimal is usually dominated by the host star during the majority of its (post) main sequence life. The direction of gravitational attraction is parallel to the planetesimal-star separation vector. As a result, there is no net torque on the planetesimal ($\bar{r}\times \bar{F}=0$), and its specific angular momentum $J$ is conserved:

\begin{equation}
J=\sqrt{GM_{{*,0}}a_0(1-{e_0}^2)}=\sqrt{GM_*(t) a(t) (1-{e(t)}^2)},
\end{equation}

\noindent where $a_0$, $e_0$ and $M_{\mathit{*,0}}$ are the initial orbital semi-major axis, eccentricity and stellar mass, respectively, and $a(t)$, $e(t)$ and $M_*(t)$ are the corresponding values at a later time $t$. 

For close-in rocky bodies, the mass loss timescale far exceeds the orbital timescale of planetesimals. In this case, mass loss is adiabatic (Appendix \ref{adiabatic}), without inducing any eccentricity variation \citep{Veras_2011}. Consequently, conservation of $J$ can be simplified to:

\begin{equation}
\label{mass loss equation}
\begin{aligned}
&a(t)=a_0 \frac{M_{*,0}}{M_*(t)},\\
&\dot a=-\frac{\dot {M_*}}{M_*}a.
\end{aligned}
\end{equation}

\subsubsection{Expanding envelope} 
\label{envelope}

During the thermally pulsing asymptotic giant branch (TPAGB), the stellar envelope of a low/intermediate-mass star expands to several AUs. Meanwhile, drastic mass loss of the TPAGB star leads to the formation of a circumstellar envelope. As a result, close-in planetesimals directly interact with stellar materials, spiralling inwards and losing masses.

First, environmental gases/dust exert ram pressure on the moving planetesimal. If ram pressure exceeds the binding energy of the body per unit volume, disruption (break up of planetesimal) would occur. The disruption threshold can be approximated as \citep{Jia_2018}:

\begin{equation}
\label{disruption equation}
 \rho_{\mathit{ext}}|\bar{v}|^2\sim\frac{GM_p\rho_p}{R_p},   
\end{equation}

\noindent where $\bar{v}$ is the velocity of the planetesimal relative to the environmental materials, and $\rho_{\mathit{ext}}$ is the density of the environment (stellar/circumstellar envelope):

\begin{equation}
\label{environmental density}
\begin{aligned}
\rho_{ext}(r)=
    \begin{cases}
        \rho_*(r) & 0<r\leq R_*\\
        -\frac{\dot M_*}{4 \pi r^2 v_{\mathit{sw}}} & r>R_*
    \end{cases}
.
\end{aligned}
\end{equation}

\noindent where $v_{\mathit{sw}}$ is the radial speed of isotropic stellar wind driven by radiation pressure, $v_{\mathit{sw}}\sim-\frac{L_*}{c\dot M_*}$, $c$ is the speed of light. The left hand side of Equation \ref{disruption equation} represents ram pressure ($P_{\mathit{ram}}$) exerted on the planetesimal and the right hand side is the gravitational binding energy ($E_b$) of the planetesimal per unit volume. For primitive (uniform composition) bodies with $M_p \propto R_p^3$, as $P_r$ is independent of $R_p$ and $E_b$ scales as $R_p^2$, disruption of the planetesimal is possibly catastrophic.

Second, motion relative to environmental gases/dust leads to friction and loss of planetesimal's angular momentum, causing inward spiraling \citep{2009ApJ...705L..81V}. The orbital speed of a planetesimal that may enter the stellar envelope during pulsations ($\sim$ 10\,km/s) far exceeds the planetesimal's escape velocity ($\sim$ 100\,m/s), indicating that the drag acting on the planetesimal is always in the hydrodynamic regime, and that gravitational focusing and accretion of environmental gases/dust onto the planetesimal is negligible. The hydrodynamic drag force can be expressed as \citep{2016MNRAS.458..832S,Jia_2018,2018ApJ...853L...1M,2023ApJ...950..128O}:

\begin{equation}
\label{friction equation}
 \bar{f}=-\frac{1}{2}C_d\rho_{\mathit{ext}}\pi {R_p}^2\bar{v}|\bar{v}|,
\end{equation}

\noindent  where $C_d$ is the drag coefficient of the planetesimal, which is a function of Mach number and Reynolds number, and approaches a constant for large relative speed $v$ (appropriate for our case) \citep{2023ApJ...950..128O}. We choose $C_d=1$ and 0.5 to investigate the qualitative behaviour of the planetesimal's inward spiralling.

We track the orbital evolution of the planetesimal throughout TPAGB by solving the equation of motion numerically with the additional friction force term in polar coordinates $(r,\phi)$, and consider that the body is lost once its disruption threshold (Equation \ref{disruption equation}) is met:

\begin{equation}
\label{equation of motion}
(\ddot r-r\dot\phi^2)\hat{r}+(2\dot r \dot\phi+r\ddot \phi)\hat{\phi}=-\frac{GM_*(<r)}{r^2}\hat{r}+\frac{\bar{f}}{M_p}.
\end{equation}

\noindent If the planetesimal's equation of motion can be decoupled
to radial and tangential components, assuming that the motion of the envelope materials during thermal pulsations is dominated by the radial component ($\bar v= (\dot{r}-v_{\mathit{ext}}) \hat{r}+r\dot{\phi} \hat{\phi}$), then:

\begin{equation}
\begin{aligned}
&\ddot r-r\dot\phi^2=\\&-\frac{GM_*(<r)}{r^2}-\frac{1}{2M_p}C_d\rho_{\mathit{ext}}(\dot r-v_{\mathit{ext}})\sqrt{(\dot r -v_{\mathit{ext}})^2+r^2\dot \phi^2},\\
&2\dot r \dot\phi+r\ddot \phi=-\frac{1}{2M_p}C_d\rho_{\mathit{ext}}r\dot \phi\sqrt{(\dot r -v_{\mathit{ext}})^2+r^2\dot \phi^2}.
\end{aligned}
\end{equation}

We additionally include the mass loss/size shrinking of the planetesimal in collisions with gases/dust (non-thermal ablation), which can be approximated as \citep{Jia_2018}:

\begin{equation}
\begin{aligned}
&\dot{M}_p\sim -2R_p^2 \rho_{\mathit{ext}}^{\frac{3}{2}} \rho_p^{-\frac{1}{2}}v,\\
&\dot{R}_p=\frac{\dot{M}_p}{4\pi R_p^2 \rho_p}.
\end{aligned}
\end{equation}

\noindent We neglect thermal ablation (sublimation) as each thermal pulse only lasts for $\sim 100$\,yr, corresponding to a diffusion length scale of $\lesssim100$\,m.

\subsection{Thermal evolution}
\label{thermal evolution}

The key to assessing the regions where a thermal process occurs is the interior temperature profile of the planetesimal. This is tracked during the evolution of the star by the radial thermal diffusion equation \citep{https://doi.org/10.1029/1998RG900006,2010AJ....140.1129J}, where we assume that spherical symmetry is preserved. When the temperature at a nodal point reaches the critical point, $T_{\mathit{crit}}$, the corresponding thermal process is assumed to occur instantaneously (see Appendix \ref{sinking} for differentiation timescale). We make the simplification that thermal and physical properties of the planetesimal remain constant. Whilst a convenient simplification for the modeling it is also the case that heat only transports via conduction as convection in 1-D is suppressed by thermal inversion (strong irradiation on the surface). 

Based on these assumptions, the radial thermal diffusion equation can be expressed as:

\begin{equation}\label{thermal diffusion equation}
\frac{\partial T}{\partial t}=\frac{1}{x^2} \frac{\partial }{\partial x}\left(\alpha_d x^2\frac{\partial T }{\partial x}\right),
\end{equation}
\noindent where $x$ is the radial coordinate, thermal diffusivity $\alpha_d=\frac{\kappa}{\rho c_p}$ is a function of thermal conductivity, $\kappa$, bulk density $\rho$ and specific heat capacity $c_p$ of the body. We choose similar thermal/physical properties as \citealp{2009E&PSL.284..144R} and \citealp{2021Sci...371..365L} (Table \ref{thermal properties table}). For planetesimals with identical radius, the dominating factor distinguishing their thermal evolution is $\alpha_d$. The sensitivity of our results to $\alpha_d$ is discussed in Section \ref{thermal discussion 2}.

\begin{table}
\begin{center}
\begin{tabular}{|c|c|c|c| } 
 \hline
 $\kappa\,(\mathrm{W/(m \cdot K)})$ &$\rho\,(\mathrm{kg/m^3})$&$c_p\,(\mathrm{J/(K\cdot kg)})$& $\alpha_d\,(\mathrm{m^2/s})$\\ 
 \hline
3 &3411.6& 1000 &$\rm8.8\times 10^{-7}$\\ 
\hline
\end{tabular}
\end{center}
\caption{Constant thermal properties of sample planetesimals.}
\label{thermal properties table}
\end{table}

We solve Equation \ref{thermal diffusion equation} numerically by implicit difference method to ensure the stability and accuracy of the long-timescale numerical integration \citep{gerya_2019}:

\begin{equation}
\begin{aligned}
&\frac{T_{i}^{n+1}-T_{i}^{n}}{\Delta t}=\\& \frac{\alpha_d}{x_i^2}\left(x_i^2\frac{T_{i+1}^{n+1}-2T_{i}^{n+1}+T_{i-1}^{n+1}}{\Delta x^2}+2x_i\frac{T_{i+1}^{n+1}-T_{i-1}^{n+1}}{2\Delta x} \right),\\&T_{i}^{n}=-\frac{\alpha_d\Delta t}{x_i^2}\left(\frac{x_i^2}{\Delta x^2}-\frac{x_i}{\Delta x}\right)T_{i-1}^{n+1}+\left(\frac{2\alpha_d\Delta t}{\Delta x^2}+1\right)T_{i}^{n+1}\\&-\frac{\alpha_d\Delta t}{x_i^2}\left(\frac{x_i^2}{\Delta x^2}+\frac{x_i}{\Delta x}\right)T_{i+1}^{n+1},
\end{aligned}
\end{equation}

\noindent where $i$ represents the number of radial steps and $n$ represents the number of time steps. We assign $10^4$ nodal points to a planetesimal's interior, and $10^5$ time steps to a planetesimal's thermal evolution during each stellar evolutionary phase (main sequence, red giant branch, core helium burning and asymptotic giant branch).

In practice, thermal properties are pressure and temperature dependent. Furthermore, heating in the body can be asymmetric, leading to convection in 3D. These factors may be important in thermal processing of a planetesimal and we discuss further the corresponding consequences in Section \ref{thermal discussion 2}.

\subsubsection{Surface equilibrium temperature}
\label{surface equilibrium temperature}

The boundary condition of Equation \ref{thermal diffusion equation} is the surface equilibrium temperature of the planetesimal, a quantification for the strength of irradiation. The surface temperature acts as the driving force of heat penetration into the interior of the planetesimal. We simplify the problem by assuming circular orbits for all planetesimals, but will discuss the effect of eccentricity by introducing time-averaged equilibrium temperature over one orbital period in Section \ref{planetesimal distribution} \citep{M_ndez_2017}. The surface equilibrium temperature of a planetesimal on a circular orbit has the form:

\begin{equation}
\label{equilibrium temperature equation}
\begin{aligned}
T_{\mathit{eq}}(r)=\left(\frac{L_*(t)(1-A_B)}{16\pi\epsilon\sigma \beta_r r(t)^2}\right)^{\frac{1}{4}},
\end{aligned}
\end{equation}

\noindent where $r$ is the distance of the planetesimal from the host star, equivalent to semi-major axis $a$ for circular orbits, $L_*$ is the stellar luminosity, $A_B$ is the bond albedo of the planetesimal, $\epsilon$ is infrared emissivity ($\epsilon \approx 1$), $\sigma$ is Stefan-Boltzmann constant, $\beta_r$ is the fraction of re-radiation ($\beta_r$ ranges from 0.5 for a tidally locked body, to 1 for a fast rotator). For simplicity, we choose $\beta_r=1$ to preserve spherical symmetry. 

\subsection{Differentiation fraction}
\label{differentiation fraction}

Based on the temperature profile evolution, We quantify the degree of differentiation in each planetesimal by the fraction of volume ($f_V$) that could have melted (had its temperature increase above $T_{\mathit{crit}}$) during the giant branches. We then repeat this process for a population of bodies with different sizes and initial semi-major axis. Noting that planetesimals are not uniformly distributed in planetesimal size ($R_p$)--initial semi-major axis ($a_0$) space, we further introduce the distribution of planetesimals $R_p$--$a_0$ space, in order to quantify the degree of differentiation of the planetesimal population.

We consider that the white dwarfs accrete rocky planetesimals that started with semi-major axes between $a_{crit}$ (critical orbit within which planetesimals end up being engulfed by the host star) and 10\,AU, with the total mass of planetesimals lying between $a_0$ and $a_0+da_0$ to be given by $\frac{\partial M_{\mathit{total}}}{\partial a_0}\propto a_0^{-\beta}$, where $\beta$ is chosen to be $\frac{1}{2}$ based on Minimum Mass Solar Nebula model (MMSN) \citep{1981PThPS..70...35H,2009ApJ...698..606C}. 

We further assume that observable white dwarf pollutants mainly come from rocky planetesimals with radii between 10 and 100\ km, with the number of planetesimals between $R_p$ and $R_p+dR_p$ to be given by $\frac{\partial N}{\partial R_p}\propto R_p^{-\alpha}$, where $\alpha$ is chosen to be 4 based on the distribution of minor objects in the Solar system \citep{2001AJ....122.2749I,2013AJ....146...36S}. The size distribution can be transformed to $ \frac{\partial M_{\mathit{total}}}{\partial R_p}\propto\frac{\partial N}{\partial R_p}M_p(R_p)\propto R_p^{3-\alpha}$. We combine spatial and size distributions of planetesimals and express the joint distribution as:

\begin{equation}
\label{distribution}
\begin{aligned}
\frac{\partial^2 M_{\mathit{total}}}{\partial R_p \partial a_0}\propto R_p^{3-\alpha}a_0^{-\beta}.
\end{aligned}
\end{equation}

The choice of power law indexes above may not apply generally in other systems. We discuss the effect of these indexes in Section \ref{planetesimal distribution}.

To assess the degree of differentiation in a system based on the joint distribution of planetesimals, we consider two scenarios. In the first, $f_{\mathit{Mm}}$, the ratio of total mass of material whose temperature exceed $T_{\mathit{crit}}$ to the total mass of planetesimals accreted onto the white dwarf is considered, regardless of whether planetesimals fully differentiate:

\begin{equation}
\label{melted region}
\begin{aligned}
f_{\mathit{Mm}}=\frac{\int\int f_V(R_p,a_0)  R_p^{3-\alpha}a_0^{-\beta}dR_pda_0}{\int\int R_p^{3-\alpha}a_0^{-\beta}dR_pda_0},
\end{aligned}
\end{equation}

\noindent where $f_V(R_p,a_0)$ is the volume fraction of differentiation for a planetesimal of radius $R_p$ with initial semi-major axis $a_0$.

In the second, $f_{\mathit{Mp}}$, only those planetesimals that melt $ > 95\%$ of their volume (defined as fully differentiated/iron core formation) are considered, under the assumption that the processes leading to the identification of core or mantle-rich material in white dwarf atmospheres require the formation of an iron core in the parent body:

\begin{equation}
\label{melted body over 95}
\begin{aligned}
f_{\mathit{Mp}}=\frac{\int\int F_V[f_V(R_p,a_0)]  R_p^{3-\alpha}a_0^{-\beta}dR_pda_0}{\int\int R_p^{3-\alpha}a_0^{-\beta}dR_pda_0},
\end{aligned}
\end{equation}

\noindent where $F_V[f_V]$ is of the form:

\begin{equation}
F_V[f_V]=
\begin{cases}
        1 & f_V\geq y\%\\
        0 & f_V < y\%
    \end{cases}
.
\end{equation}

The applied integration limits to Equation \ref{melted region} and \ref{melted body over 95} are summarized in Table \ref{integration limit table}.

\begin{table}
\begin{center}
\begin{tabular}{|c|c|c|c| } 
 \hline
$a_{0,\mathit{min}}$ &$a_{0,\mathit{max}}$&${R}_{p,\mathit{min}}$&  ${R}_{p,\mathit{max}}$\\ 
 \hline
$a_{crit}$ &10\,AU& 10\,km &100\,km\\ 
\hline
\end{tabular}
\end{center}
\caption{Integration limits of sample planetesimals' initial semi-major axis ($a_0$) and size ($R_p$) for Equation \ref{melted region} and \ref{melted body over 95}.}
\label{integration limit table}
\end{table}

\section{Results}
\label{results}

\subsection{Stellar evolutionary tracks}
\label{stellar evolutionary tracks}

\begin{figure}
\begin{center}
\includegraphics[width=0.45\textwidth]{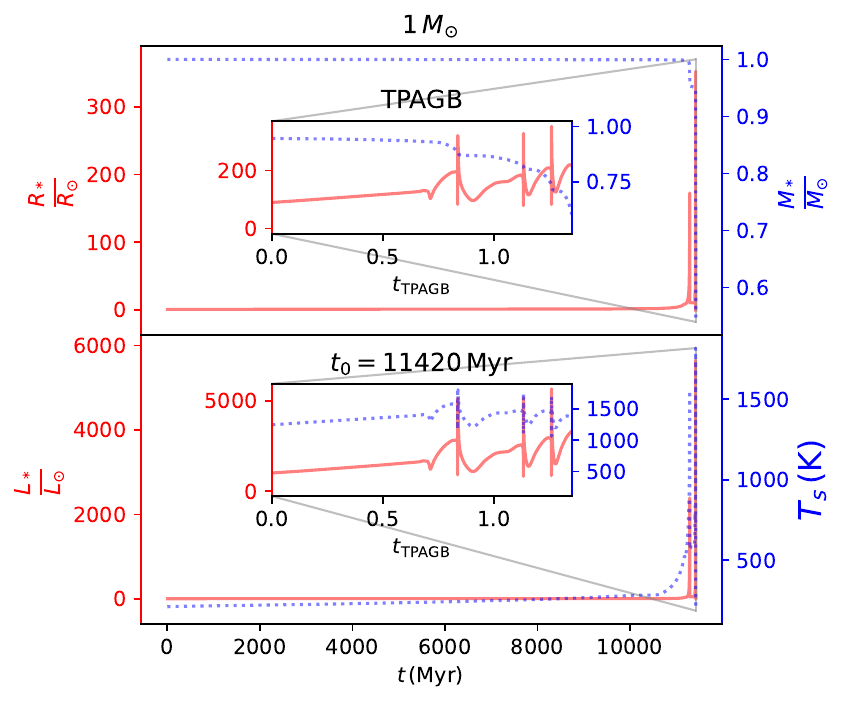}
\includegraphics[width=0.45\textwidth]{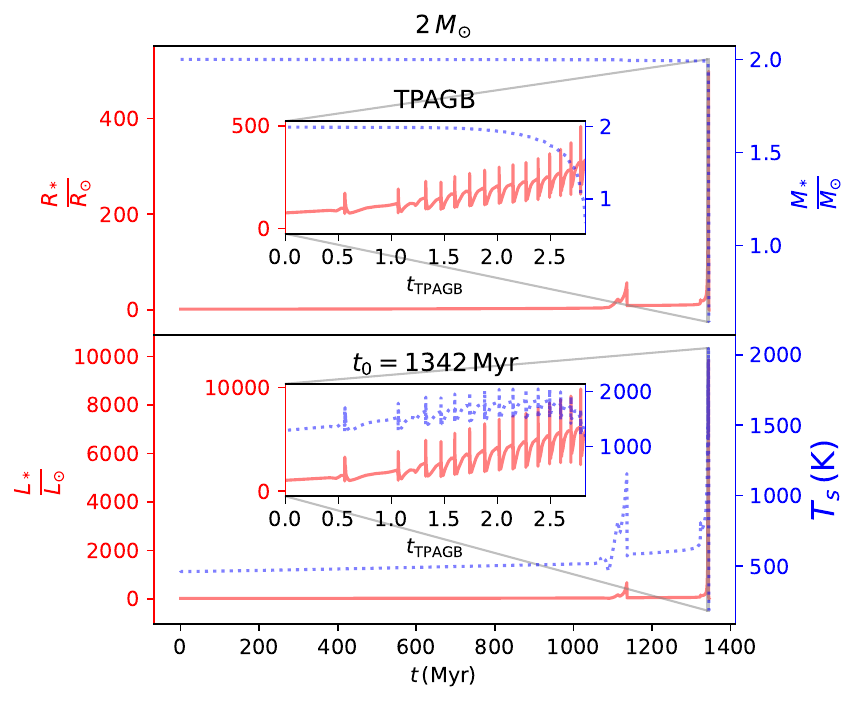}   
\caption{Evolution of stars with $M_*=1\,M_{\odot}$ and 2\,$M_{\odot}$ and the resultant equilibrium temperature for planetesimals with $a_0=1.5$\,AU and $e=0$. The zoom-in plots correspond to thermally pulsing asymptotic giant branch (TPAGB) with the origin of the time axes marks the start of TPAGB phase. The age of the star at the start of of its TPAGB phase ($t_0$) is added.}
\label{stellar evolution plot}
\end{center}
\end{figure}

\begin{table}
\begin{center}
\begin{tabular}{ |c c|c|c|c| } 
 \hline
 & &end MS &tip RGB& tip AGB \\ 
 \hline
$L_*\,(L_{\odot})$ &$1\,M_{\odot}$ & 2 & 2273 &5657\\ 
&$2\,M_{\odot}$ & 36 & 654& 9858\\ 
 \hline
 $R_*\,(R_{\odot})$ &$1\,M_{\odot}$ & 1.5 & 166.5&352.3\\ 
&$2\,M_{\odot}$ & 4.1 & 56.0 &498.2\\ 
\hline
\end{tabular}
\end{center}
\caption{Comparison of the maximum stellar luminosity and radius during the main sequence, red giant branch and asymptotic giant branch between the 1 and 2 solar mass star.}
\label{stellar parameter table}
\end{table}

In this section we summarise and compare the MIST evolutionary tracks (Section \ref{stellar evolution}) of two samples, I: $1\,M_{\odot}$ low-mass star and II: $2\,M_{\odot}$ intermediate-mass star (Figure \ref{stellar evolution plot}), both of which go through AGB and end in C-O white dwarfs, while undertaking distinct evolutionary tracks near their RGB tips. 

Figure \ref{stellar evolution plot} and Table \ref{stellar parameter table} illustrate that sample I is more enhanced in size and luminosity during its RGB tip (approaching half of its tip AGB values) compared to sample II, whose tip RGB luminosity (radius) is around $1\%$ of its tip AGB value. This occurs because low-mass ($\sim 0.8$--$2\,M_{\odot}$) stars, unlike their intermediate-mass ($\sim 2$--$8 \,M_{\odot}$) counterparts, lack gravitational pressure in their cores and start helium burning with helium flash (runaway fusion in the contracted degenerate cores), until thermal pressure dominates, boosting their sizes and luminosity significantly at the RGB tips.

Mass loss patterns for both samples are similar: rapid mass loss near the end of TPAGB. In this case, Equation \ref{mass loss equation} indicates that outward migration of planetesimals mainly occurs at the end of thermal pulsations, suppressing the growth in planetesimal surface temperature ($T_s$, Equation \ref{equilibrium temperature equation}) with stellar luminosity ($L_*$). The $T_s$ of a planetesimal initially orbiting sample II at 1.5\,AU shows clear decreasing trend near the end of TPAGB, despite the generally increasing $L_*$ (Figure \ref{stellar evolution plot}, lower panel). For the same reason, planetesimals are much more likely to escape from the envelope via stellar mass loss during late pulsations.  

\subsection{Temperature profile evolution}
\label{temperature profile evolution}

\begin{figure*}
\includegraphics[width=\textwidth]{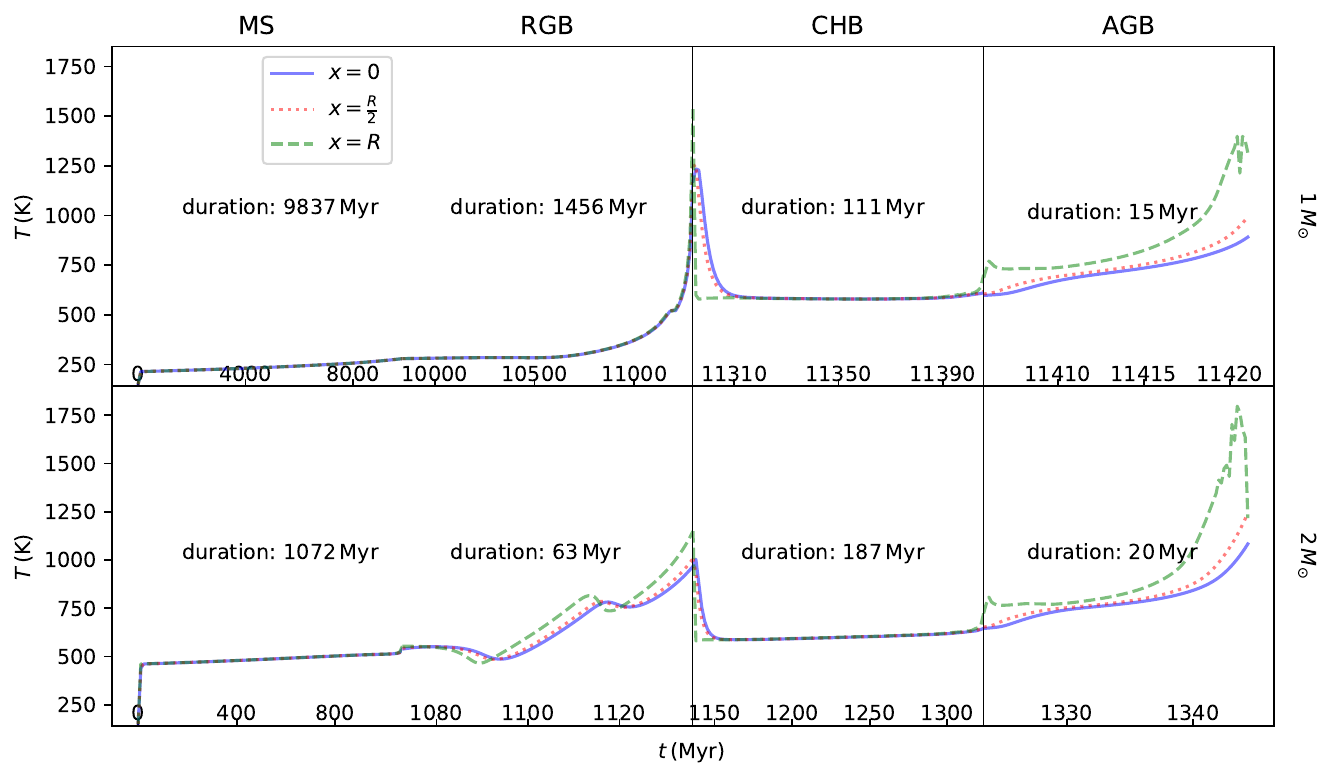}   
\includegraphics[width=\textwidth]{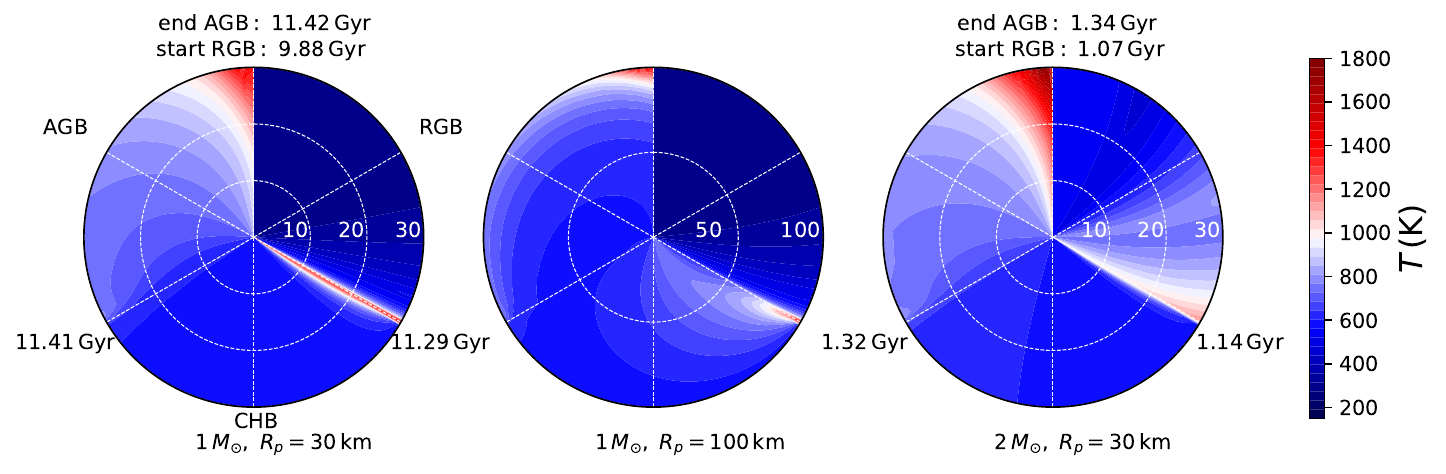}   
\caption{The upper panel shows the temperature evolution at the centre ($x=0$), half of the radius ($x=\frac{R}{2}$) and the surface ($x=R$) of a planetesimal in sample system IA and IIA (Table \ref{temperature profile evolution sample table}). The period of each evolutionary phase is normalised by its overall timescale, with the absolute time along the horizontal axis of the plot. The lower panel shows the temperature profile evolution of planetesimals in sample system IA, IB and IIA. The contour plots start from the red giant branch (RGB) of the host star, followed by the core helium burning phase (CHB) and the asymptotic giant branch (AGB), with the age of the star increasing clock-wisely. Same normalisation as the upper panel is applied.}
\label{thermal evolution plot}
\end{figure*}

As is described in Section \ref{thermal evolution}, the time evolution of planetesimal's interior temperature profile is calculated based on the stellar irradiation, quantified by $T_s$ (Equation \ref{equilibrium temperature equation}). We show the thermal evolution of 3 sample systems (listed in Table \ref{temperature profile evolution sample table}) in Figure \ref{thermal evolution plot}.

\begin{table}
\begin{center}
\begin{tabular}{ |c |c|c|c|c| } 
 \hline
  Sample&$M_*\,(M_{\odot})$&Sub-sample&$R_p$\,(km)& $a_0$\,(AU)  \\
 \hline
 I&1&IA&30 & 1.5 \\ 
&&IB&100&1.5\\ 
 \hline
II& 2&IIA &30&1.5\\
\hline
\end{tabular}
\end{center}
\caption{List of the parameters of 3 sample systems in Figure \ref{thermal evolution plot}.}
\label{temperature profile evolution sample table}
\end{table}

In all 3 samples, the temperature evolution in the interior of planetesimals lags behind that of its surface temperature ($T_s$). The time delay of a layer's temperature relative to $T_s$ can be estimated by the diffusion time scale ($t_{\mathit{diff}}=\frac{d_{\mathit{diff}}^2}{\alpha_d}$) corresponding to the depth of this layer and is especially relevant for rapid $T_s$ variations, e.g., during the tip RGB and TPAGB. For instance, thermal pulsations lasting $\sim$ 100--1000\,yr correspond to a diffusion length scale $d_{\mathit{diff}}\sim$ 100\,m, thus only affecting a thin surface layer of this size.

The interior temperature of the planetesimals in sample IA and IIA rises to the corresponding equilibrium temperature within a small fraction of stellar main sequence life, since the diffusion timescale ($\tau_{\mathit{diff}}$) of 30\,km is around 30\,Myr, much shorter compared to the main-sequence lifetime of the host stars. The initial temperature profile of planetesimals are unimportant unless the host star is more massive than 3\,$M_{\odot}$ and $R_p$ exceeds 100\,km (at which point, the diffusion length scale corresponds to the main sequence lifetime of these shorter-lived stars satisfies: $d_{\mathit{diff}}(\tau_{\mathit{main\ sequence}})\lesssim R_p$). This is supported by the uniform temperature profile of these 3 samples at the start of RGB (Figure \ref{thermal evolution plot}, lower panel).

Most of the thermal processes in planetesimals of sample IA and IB occur at the tip RGB, while AGB heating in IA and IB only penetrates layers of depths $\sim$ 10\,km because of its short duration. On the contrary in sample IIA, AGB heating is much more significant than its RGB counterpart. This is consistent with the results in Section \ref{stellar evolutionary tracks}: the enhanced RGB luminosity of low-mass stars due to helium flash increases their ability to heat planetesimals during the RGB stage, compared with their AGB stage, whereas intermediate-mass stars do not go through the He flash, and therefore their planetesimal heating is dominated by the AGB phase.

When exposed to intense stellar irradiation (as in the case during the giant branches), a planetesimal's interior temperature decreases inwards (thermal inversion). Thermal processing requiring strong irradiation, e.g., differentiation, starts at the exterior of planetesimals. This thermal inversion ceases at the start of CHB phase, when the stellar luminosity drops before rising up again. At this point, the interior temperature of a planetesimal decreases outwards at its outer layer, preserving longer if the body is larger (sample IB). 

\subsection{Envelope entry}

\begin{figure}
\includegraphics[width=0.45\textwidth]{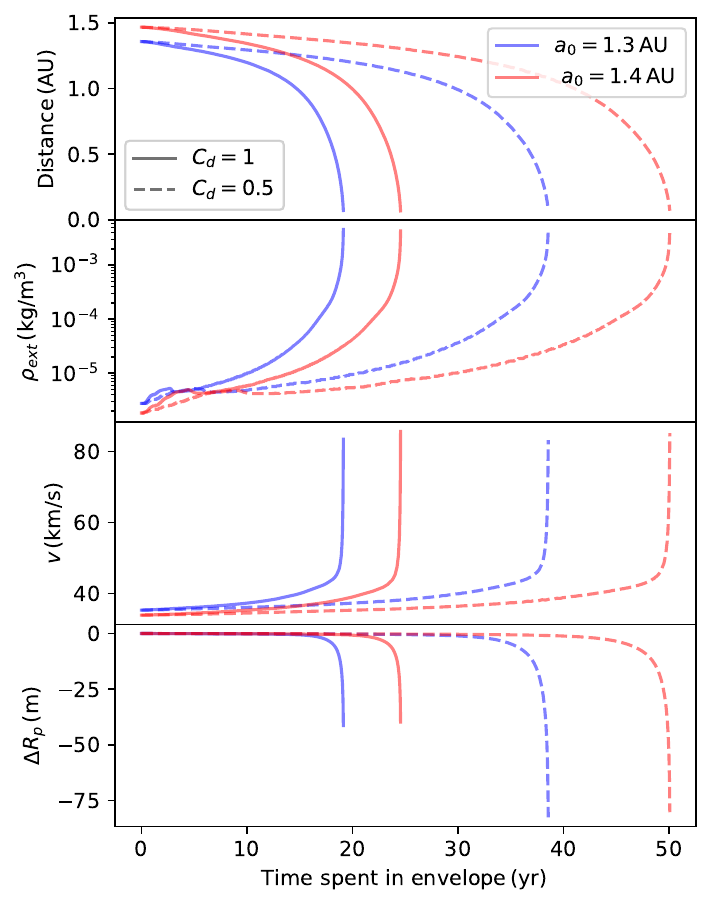}   
\caption{Orbital decay and ablation of planetesimals with different initial semi-major axis ($a_0$) and drag coefficient ($C_d$) inside the stellar envelope. $M_*=2\,M_{\odot}$, $R_p=100$\,km, $e=0$.}
\label{envelope spiral in plot}
\end{figure}

In this section we track the orbital evolution of planetesimals that eventually enter the stellar envelope during its TPAGB, based on planetesimal's equation of motion (Section \ref{envelope}). Our simulation verifies the planetesimal's orbital evolution is still dominated by outward migration from stellar mass loss, rather than stellar wind drag before envelope entry (see Appendix \ref{stellar wind}). We present the orbital evolution and mass loss for 4 sample planetesimals of $R_p=100$\,km after entering the envelope of a 2\,$M_{\odot}$ star, with $a_0=1.3$ and 1.4\,AU, and $C_d=1$ and 0.5, respectively, in Figure \ref{envelope spiral in plot}.

Planetesimals in all four systems considered spiral inwards to their disruption limits (Equation \ref{disruption equation}) within 50 yr. According to Equation \ref{friction equation}, deceleration due to drag force scales as $\frac{R_p^2}{M_p}\propto\frac{1}{R_p}$, indicating that smaller planetesimals experience a stronger deceleration from drag and spiral in faster. Therefore, planetesimals with $R_p\lesssim100$\,km cannot survive in the stellar envelope when ignoring other effects (e.g., interactions with the planetary system). Meanwhile, much larger planetesimals/planets, for instance, with $R_p\sim 1000$\,km feels less drag effect ($\lesssim 10 \%$ decay in $a$ during the first envelope entry) and may have enough energy to (partially) unbind the envelope.

Planetesimal mass loss due to ablation is negligible in all 4 samples. Ablation mainly occurs when the planetesimal is close to its disruption limit, where both environmental density and planetesimal's speed relative to the stellar envelope increase rapidly. 

\subsection{Thermal processing inside planetesimals}
\label{thermal processing inside planetesimals}

In this section we present the degree of thermal processing/differentiation in planetesimal radius ($R_p$)--initial semi-major axis ($a_0$) space based on thermal equations described in Section \ref{thermal evolution}. Section \ref{maximum central temperature} shows the maximum temperature raised at the centre ($x=0$) and half radius ($x=\frac{R_p}{2}$) of the sample planetesimals in our parameter space. Section \ref{volume fraction of melting} demonstrates the volume fraction of melting ($f_V$) in this space for two temperature thresholds ($T_{crit}$), 1400\,K and 1800\,K. Based on $f_V(R_p,a_0)$, we present, in Section \ref{diff fraction}, the estimated mass fraction of differentiated samples according to two definitions, $f_{\mathit{Mm}}$ in Equation \ref{melted region} and $f_{\mathit{Mp}}$ in \ref{melted body over 95}.

\subsubsection{Maximum central temperature}
\label{maximum central temperature}

\begin{figure*}
\includegraphics[width=\textwidth]{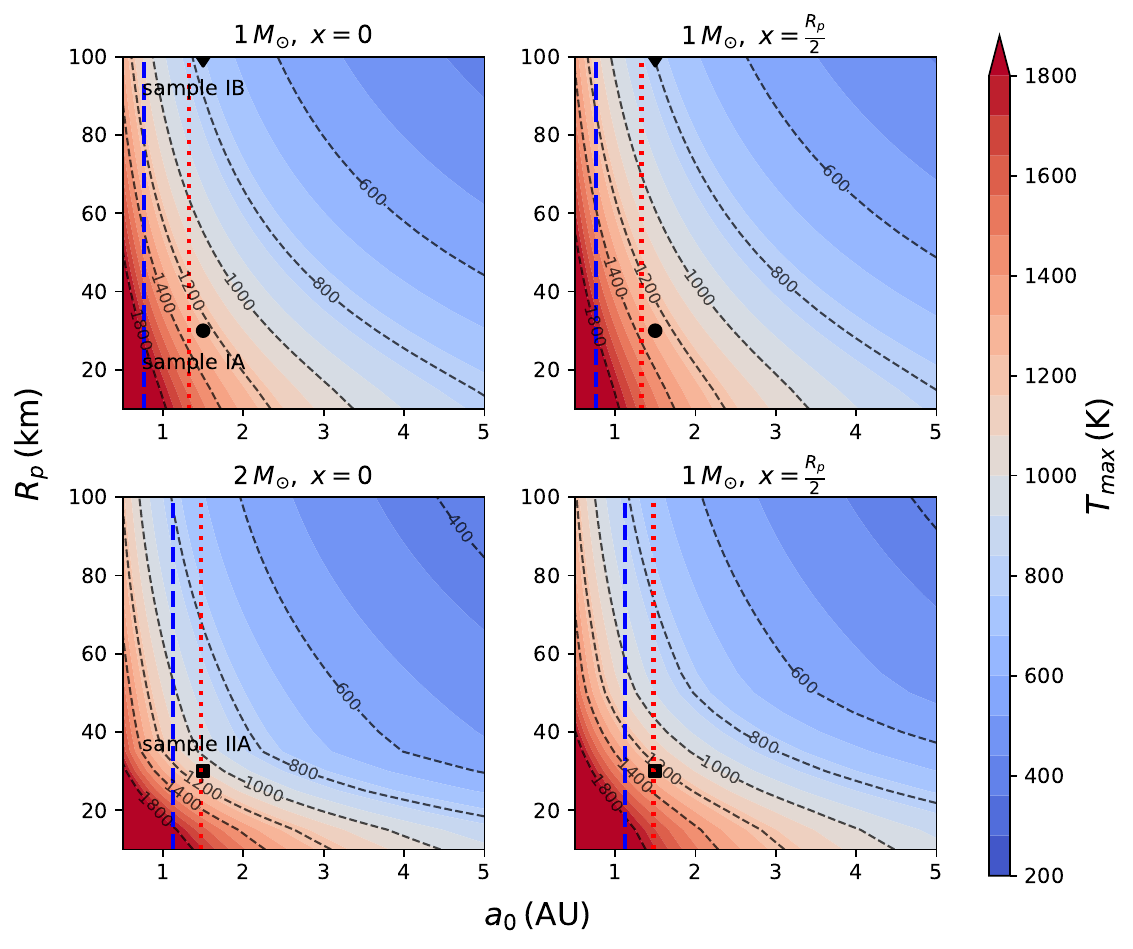}  
\caption{The maximum temperature at the centre ($x=0$) and half of the radius ($x=\frac{R_p}{2}$) of the planetesimal, throughout its thermal history under various circumstances. The blue (red) dashed (dotted) line corresponds to the critical orbit after (before) smoothing pulsations. The positions of sample IA (black circle) IB (triangle) and IIA (square) in Figure \ref{thermal evolution plot} are added.}
\label{central T plot}
\end{figure*}

We present the maximum temperature at $x=0$ and $\frac{R_p}{2}$ in planetesimals of two sample systems, I: 1\,$M_{\odot}$ and II: 2\,$M_{\odot}$ host star, in Figure \ref{central T plot}. This figure shows that any thermal process triggered by stellar evolution take place preferentially in smaller-sized bodies closer to the host star, with its $R_p$ and $a_0$ dependence determined by the corresponding temperature threshold. 

As is shown in the left panels of Figure \ref{central T plot}, all completely melted planetesimals (which satisfy $T(x=0)\geq T_{crit}=1800$\,K) later enter the stellar envelope during the thermally pulsing AGB (fall to the left of the red dotted line). Considering the uncertainties in modeling thermal pulsations of AGB stars, we smooth out the oscillations of stellar radius and compute the corresponding engulfment limit, represented by the blue dashed lines in Figure \ref{central T plot} (see Section \ref{pulsation discussion}). This new engulfment limit leaves a narrow triangle-like parameter space of completely melted planetesimals that survive the AGB phase of the host star (sample I: $0.8\,\mathrm{AU} \lesssim a_0 \lesssim 1\,\mathrm{AU}$, $ R_p \lesssim 28$\,km, II: $1\,\mathrm{AU}\lesssim a_0 \lesssim1.3\,\mathrm{AU}$, $ R_p\lesssim18$\,km). This result emphasizes the importance of thermal pulsations on the survival of the most thermally processed planetesimals.

On the other hand, if $T_{\mathit{crit}}$ is lowered to 1400\,K (see Section \ref{thermal discussion1} for the reason of this choice), some completely melted planetesimals survive the host star's thermal pulsations (sample I: $1.3\,\mathrm{AU}\lesssim a_0 \lesssim1.7\,\mathrm{AU}$, $ R_p \lesssim 28$\,km, II: $1.5\,\mathrm{AU}\lesssim a_0 \lesssim 2.2\,\mathrm{AU}$, $R_p \lesssim 22$ km). Meanwhile, there is an increasingly larger area under a lower temperature contour in Figure \ref{central T plot}. In other words, a lower $T_{\mathit{crit}}$ corresponds to a much larger $R_p$--$a_0$ space where the planetesimal is heated above this threshold over a given fraction of its volume. This result stresses the strong dependence of the degree of differentiation on $T_{\mathit{crit}}$.

The thermal processing in sample I is mainly limited by the strength of heating, whereas in sample II the thermal processing is mainly limited by the timescale of heating. Compared to sample I (upper panels of Figure \ref{central T plot}) that undergoes helium flash during its RGB, a given temperature contour in sample II (lower panels, quiet ignition of core helium burning) intersects the axes at smaller $R_p$ but larger $a_0$ (for instance, see the 1800\,K contour lines). Furthermore, the increase in $R_p$ value of each contour line--the $R_p$ axis intersections from $x=0$ (left panels of Figure \ref{central T plot}) to $x=\frac{R_p}{2}$ (right panels) is larger for sample II (lower panels), indicating that compared to sample I, the temperature gradient in sample II is steeper. These features are consistent with the facts that 1. the maximum thermal processing in sample I occurs at the RGB phase, which is much longer (lasting around 1.5\,Gyr, corresponding to $d_{\mathit{diff}}\sim 100$\,km) with much smaller average luminosity gradient ($1.7\,L_{\odot}\mathrm{/Myr}$) compared to the AGB life of sample II (20\,Myr, $d_{\mathit{diff}}\sim10$\,km, $272.3\,L_{\odot}\mathrm{/Myr}$, when planetesimals undergo the maximum heating), and 2. the tip AGB of sample II is around five times as luminous as the tip RGB of sample I. 

\subsubsection{Volume fraction of melting}
\label{volume fraction of melting}

\begin{figure*}
\includegraphics[width=\textwidth]{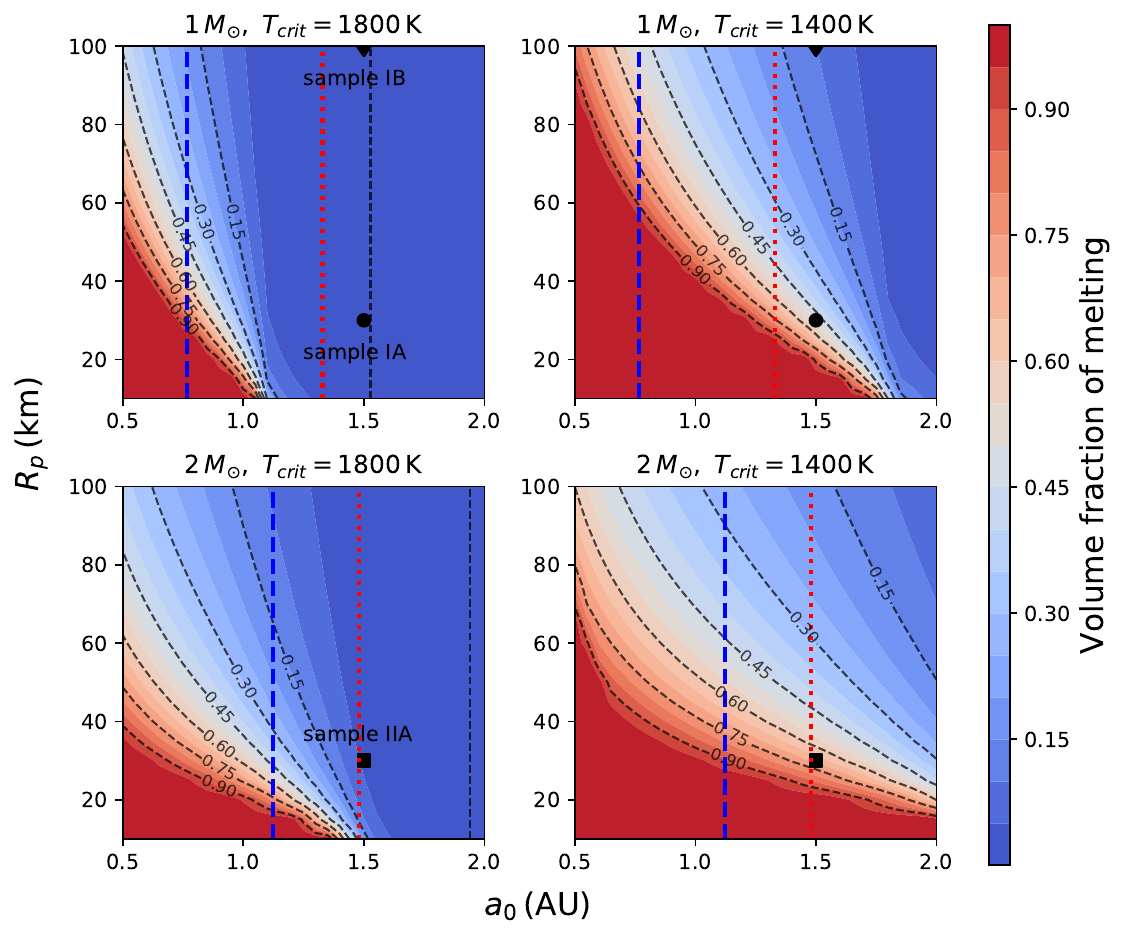}   
\caption{Volume fraction that is once melted ($f_V$) for planetesimals of different sizes and initial positions. The blue (red) dashed (dotted) lines correspond to the critical orbit after (before) smoothing stellar thermal pulsations. The black dashed vertical line corresponds to the semi-major axis beyond which the maximum surface temperature of planetesimals drop below the given melting threshold. The positions of sample IA (black circle) IB (triangle) and IIA (square) in Figure \ref{thermal evolution plot} are added.}
\label{volume fraction of diff plot}
\end{figure*}

We present the fraction of volume that reaches sufficient temperatures for melting at any point during its evolution of each planetesimal in $R_p$--$a_0$ space ($f_V(R_p,a_0)$) for two sample systems, I: $1\,M_{\odot}$ and II: $2\,M_{\odot}$ host star considering two $T_{\mathit{crit}}$, 1400\,K (right panels) and 1800\,K (left panels) in Figure \ref{volume fraction of diff plot}.

The similarities between Figure \ref{central T plot} and Figure \ref{volume fraction of diff plot} are summarized below:

\begin{itemize}
    \item Smaller planetesimals closer to the star undergo larger-scale melting 
    \item All planetesimals where more than $10\%$ by volume reaches temperatures above 1800\,K are engulfed
    \item Compared to 1800\,K, the fraction of planetesimals where a portion of their volume reaches above 1400\,K increases significantly
\end{itemize}

With the decrease in $R_p$, the contours of Figure \ref{volume fraction of diff plot} converge to the critical value of $a_0$ (black dashed vertical lines in the left panels), beyond which the surface temperature of planetesimals can never reach $T_{crit}$. This convergence results from the decrease of the duration when $T_s>T_{crit}$ with the increase in $a_0$.

\subsubsection{The degree of differentiation}
\label{diff fraction}

\begin{figure}
\includegraphics[width=0.45\textwidth]{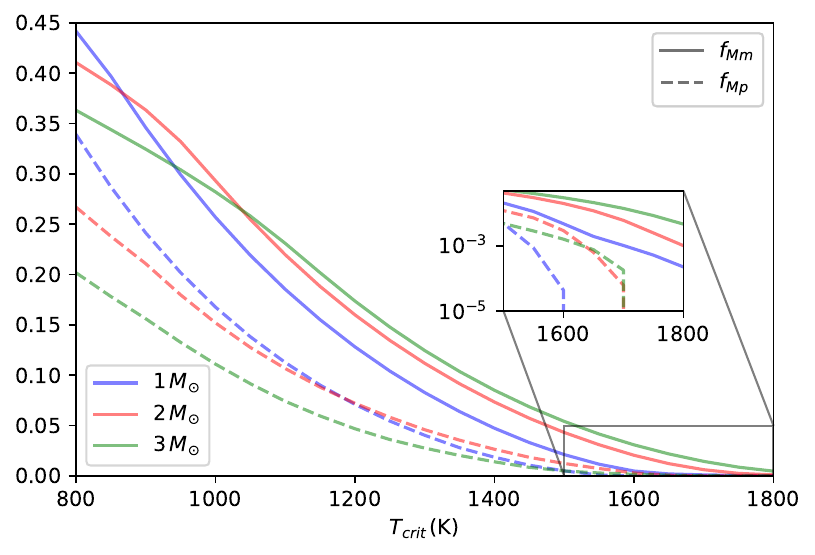}   
\caption{Mass fraction of melted regions in single planetesimals ($f_{\mathit{Mm}}$, solid lines), and mass fraction of planetesimals with over $95\%$  of the volume undertaking melting ($f_{\mathit{Mp}}$, dashed lines). We choose $\alpha=4$, $\beta=\frac{1}{2}$ in Equation \ref{melted region} and \ref{melted body over 95}.}
\label{mass fraction of diff plot}
\end{figure}

Due to the fact that observations of white dwarf pollutants may occur at any point of a white dwarf's accretion history, such that the sizes and initial positions of the accreted bodies are unconstrained, it is useful to consider the average effect of stellar irradiation on the interiors of a population of planetesimals. We quantify this average effect by the mass fraction of the planetesimal population where large-scale melting is induced by the host star's giant branches. In order to calculate this fraction, we consider a nominal population of planetesimals, orbiting within 10\,AU surviving the giant branches, with sizes between 10 and 100\,km, as described in Section \ref{differentiation fraction} with $\alpha=4$, $\beta=\frac{1}{2}$ ($\frac{\partial^2 M_{\mathit{total}}}{\partial R_p \partial a_0}\propto R_p^{-1}a_0^{-\frac{1}{2}}$). The volume fraction of melting ($f_V(R_p,a_0)$), as described in Section \ref{volume fraction of melting} is integrated over planetesimal size ($R_p$)--initial semi-major axis ($a_0$) space to get the mass fraction of melted regions ($f_{\mathit{Mm}}$, Equation \ref{melted region}), and the mass fraction of planetesimals melted over $95\%$ of the volume ($f_{\mathit{Mp}}$ with $y=95$, Equation \ref{melted body over 95}), relative to the total mass of the planetesimal population. In this section we present $f_{\mathit{Mm}}$ and $f_{\mathit{Mp}}$ at different $T_{\mathit{crit}}$ in 3 sample systems, I: 1\,$M_{\odot}$ and II: 2\,$M_{\odot}$ and III: 3\,$M_{\odot}$ host star in Figure \ref{mass fraction of diff plot}.

Both $f_{\mathit{Mm}}$ and $f_{\mathit{Mp}}$ in all 3 samples increase rapidly with the decrease in the critical temperature of melting, $T_{\mathit{crit}}$. There are 2 orders of magnitude difference from 800 K to 1800 K in $f_{Mm}$ for all 3 samples. $f_{\mathit{Mp}}$ always lie below the corresponding $f_{\mathit{Mm}}$, converging rapidly to 0 when $T_{\mathit{crit}}$ approaches the maximum surface temperature of planetesimals within 100 K, indicating its much stricter requirement of small $R_p$ and $a_0$ (long-lasting intense heating). For $f_{\mathit{Mm}}$ ($f_{\mathit{Mp}}$) to exceed $10\%$, $T_{\mathit{crit}}$ requires lowering to $\lesssim$ 1300\,K (1200\,K) for sample I, $\lesssim1400$\,K (1200\,K) for sample II and $\lesssim1400$\,K (1000\,K) for sample III. 

At $T_{\mathit{crit}}\gtrsim$ 1100\,K, $f_{\mathit{Mm}}$ of identical $T_{\mathit{crit}}$ decreases with stellar mass. In this region, the rapid increase in stellar luminosity with stellar mass is the dominant factor affecting the degree of differentiation. The contrary occurs at $T_{\mathit{crit}}\lesssim900$\,K, where the duration of intense irradiation becomes more important. A similar feature presents in $f_{\mathit{Mp}}$, at $T_{\mathit{crit}}\gtrsim1700$\,K and $\lesssim1200$\,K, stressing the growing importance of the timescale of intense heating for large-scale melting. 

\section{Discussion}
\label{discussion}

Our model predicts that stellar evolution may lead to differentiation in small planetesimals close to the host star. However, the size of the parameter space where large-scale melting occurs, as well as its contributions to the white dwarf pollutants will largely depend on how thermal pulsations occur and the critical temperature of melting: for differentiated planetesimals to occupy $10\%$ of the mass of the population described in Section \ref{diff fraction}, $T_{\mathit{crit}}$ must be lowered to $\sim$ 1300 K. 

In this section we discuss the main limitations in this study: stellar/planetesimal orbital evolution during TPAGB (Section \ref{pulsation discussion}), thermal evolution of planetesimals (Section \ref{thermal discussion1} and \ref{thermal discussion 2}), and planetesimal distributions (Section \ref{planetesimal distribution}), together with their possible impacts on the results. Based on the white dwarf atmospheric model, we discuss the observability of accretion of planetesimals differentiated under stellar irradiation (Section \ref{observational limit}).  

By comparing our predictions to observational data of polluted white dwarfs, we further discuss the necessity of early thermal processes such as radiogenic heating, impacts and incomplete condensation (Section \ref{implications}). 

\subsection{Limitations}
\label{limitations}

\subsubsection{Thermal pulsing asymptotic giant branch}
\label{pulsation discussion}

\begin{figure}
\includegraphics[width=0.45\textwidth]{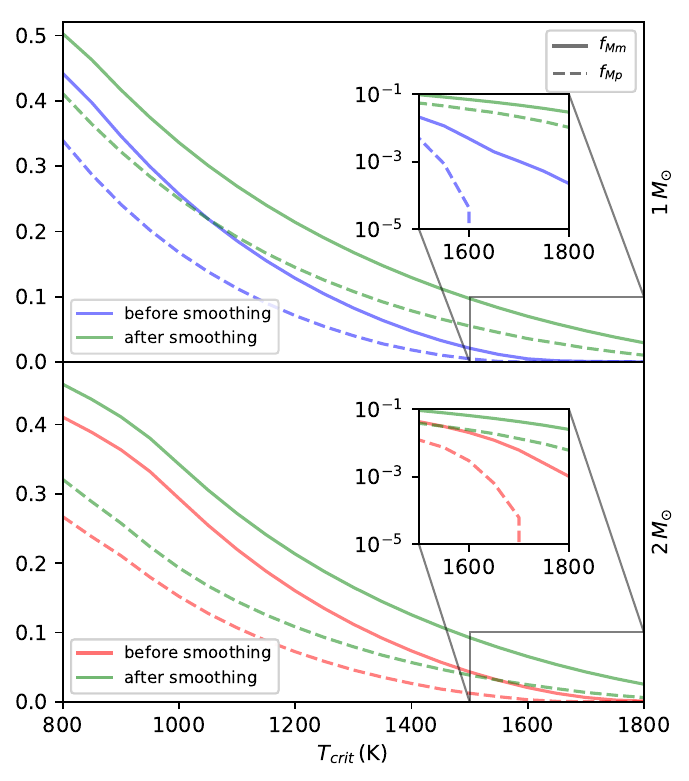}   
\caption{Comparison between $f_{\mathit{Mm}}$, mass fraction of melted regions in single planetesimals, and $f_{\mathit{Mp}}$, mass fraction of planetesimals melted over $95\%$ of the volume, relative to the total mass of the planetesimal population, in 1 (blue, upper panel) and 2 (red, lower panel) solar mass star systems, before and after smoothing out pulsations (green), with the same assumptions in Figure \ref{mass fraction of diff plot}.}
\label{mass fraction plot without pulsation}
\end{figure}

Thermal pulsations are accompanied with unstable cyclical nuclear burning at the base of He and H shells. No steady state is available because of the significantly different reaction rates of H and He burning. Powered by the intense helium flash, a convective zone forms in the He intershell, affecting stellar pulsations and dredge-ups \citep{1994sipp.book.....H,2015ApJS..219...40C}. As a result, thermal pulsations are extremely sensitive to specific assumptions of convection, for instance, convective overshooting. In reality, convection is a 3-D process, which cannot be fully simulated by MESA, a 1-D stellar evolution code \citep{2021zndo...5798242P,2023ApJS..265...15J}.

On the other hand, with the rapid mass loss of the host star during thermal pulsations, the gravitational attraction from the host star weakens, while the interactions of planetesimals with the planetary system becomes stronger. The orbital evolution of planetesimals may be dominated by resonances, and planetesimals may be scattered inwards/outwards. A massive planet can also unbind the loosely bound pulsating stellar envelope, altering the condition of engulfment and the stellar evolutionary track.

However, we have shown that the expansion of the stellar envelope during thermal pulsations, without any planetesimal--planet interaction, leads to engulfment of the most thermally processed planetesimals. For instance, Figure \ref{volume fraction of diff plot} shows that almost all planetesimals with melting covering $>15\%$ of their volumes ($f_V\gtrsim 15\%$) with $T_{\mathit{crit}}=1800$\,K are engulfed during the TPAGB. 

To account for the uncertainties in modelling thermal pulsations and the additional effect of the planetary system, we repeat the analysis in Section \ref{diff fraction} while smoothing out pulsations (similar to SSE code, \citealp{2013ascl.soft03015H}) for two sample systems, I: 1\,$M_{\odot}$ and II: 2\,$M_{\odot}$ host star, shown in Figure \ref{mass fraction plot without pulsation}. In both samples, there is $\sim 5\%$--$10\%$ increase in $f_{\mathit{Mm}}$ and $f_{\mathit{Mp}}$. This increase is especially significant at high $T_{\mathit{crit}}$, e.g., 1600\,K--1800\,K, where orders of magnitude difference are present.

Furthermore, the luminous AGB phase may lead to non-negligible Yarkovsky effect and YORP effect arising from asymmetric thermal emission of planetesimals, altering their orbital (semi-major axis and eccentricity) and spin parameters (spin angular velocity and obliquity). We simulate the maximum Yarkovsky effect and the maximum YORP effect acting on a planetesimal (see Appendix \ref{Ya appendix}). Our simulations illustrate that for our smallest planetesimals ($R_p=10$\,km ), the outward/inward Yarkovsky drift can reach $\sim 0.4$\,AU at the end of TPAGB life of the system. However, compared to the uncertainty in the planetesimal's critical semi-major axis at the end of the TPAGB phase ($\sim$ 1\,AU) arising from modelling of thermal pulsations, and accounting for the fact that the maximum Yarkovsky drift is only reached under extreme circumstances (e.g., tidal locking, maximum/minimum entries of Yarkovsky matrix), the Yarkovsky effect may only play a minor role in the survival of close-in planetesimals. On the other hand, the maximum YORP spin up has the possibility to break up the smallest planetesimals whose angular velocity already exceeds $\sim0.5\,\omega_{\mathit{break}}(=\sqrt{\frac{4\pi G\rho}{3}})$. Therefore, when accounting for the YORP effect, there may be an extra loss in the smallest and hence the most thermally processed planetesimals, resulting in fewer planetesimals accreted by the white dwarf that underwent large-scale melting induced by giant branch heating.

Another uncertainty of TPAGB is the mass loss pattern. Although an adiabatic approximation is still valid in our parameter space, mass loss could be axial instead of isotropic, such that only a small proportion of stellar luminosity contributes to the momentum of ejecting masses, leading to much slower stellar wind and denser circumstellar envelope. When exposed to the dense axial stellar wind, ablation of planetesimal becomes faster, accompanied with faster inward spiraling. As a result, although the orbital evolution of planetesimals unaffected by the axial stellar wind remain dominated by stellar mass loss, the engulfment regions, and thus the critical orbits, extend outwards for planetesimals inside the overdense stellar wind. More close-in (the most thermally processed) planetesimals are engulfed during the TPAGB, reducing the degree of thermal processing of the whole planetesimal population ending up accreted by the white dwarf.

In this study we assume a constant metallicity, 0.014 for all sample host stars, which is unable to represent the whole population of white dwarfs. For stars undergoing identical evolutionary phases, a higher metallicity corresponds to larger opacity, and hence a less compact star with more energy dissipated in the interior, resulting in stronger thermal expansion during pulsations. Meanwhile, stars with considerably distinct metallicty may undergo distinct evolutionary phases. For instance, when the metallicity increases from 0.014 to $\sim 0.05$ for a 1 solar mass star, the star no longer undergoes an AGB phase, corresponding to a scenario that more close-in planetesimals avoid engulfment with the degree of thermal processing (dominated by the RGB phase) in individual bodies remained. The degree of thermal processing averaging over all planetesimals accreted by the white dwarf hence increases.

\subsubsection{Critical point of melting}
\label{thermal discussion1}

In this work we consider a simple model in which any portion of a planetesimal heated above a critical temperature is considered to have melted. The key question is whether a single critical point is valid and if so, what the appropriate critical temperature would be. Planetesimals have mixed compositions and hence melting ranges, whose boundaries correspond to the highest temperature for solidus ($T_{\mathit{solidus}}$, no melting) and the lowest temperature for liquidus ($T_{\mathit{liquidus}}$ complete melting), instead of a single melting point marking the critical transition from solid to liquid. However, we can define a critical point of melting, as the transition from solid-like to liquid-like behaviours, which occurs typically around a liquid fraction of 40\% \citep{2005GeoRL..3222308C,2007E&PSL.264..402C}. This critical liquid fraction may be reached at different temperatures (within the melting range) in different regions depending on the exact composition, local gravity, etc. However, given the other uncertainties in this model, using a single temperature provides a reasonable qualitative behaviour. 

The uncertainty in the critical point ($T_{\mathit{crit}}$) lies in both the poorly constrained melting range ($T_{\mathit{solidus}}$--$T_{\mathit{liquidus}}$) and the position of $T_{\mathit{crit}}$ within this melting range: the degree of melting required for efficient melt migration.

According to the model in \citealp{2023A&A...671A..74J}, the pressure dependent melting range boundaries, $T_{\mathit{solidus}}$ and $T_{\mathit{liquidus}}$ for planetary silicates can be estimated as:

\begin{equation}
T_{\mathit{melt}}=T_0\left(1+\frac{P}{P_0}\right)^q,
\end{equation}

\noindent where $T_0\sim 1661$\,K, $P_0\sim1.336$\,GPa, $q\sim0.134$ for $T_{\mathit{solidus}}$ and $T_0\sim 1982$\,K, $P_0\sim6.594$\,GPa, $q\sim 0.186$ for $T_{\mathit{liquidus}}$. For the largest primitive planetesimal in our parameter space ($R_p$=100 km), there is only an increase of $\approx3$ (1)\,K in $T_{\mathit{solidus}}$ ($T_{\mathit{liquidus}}$) from the surface to the centre of the body because of pressure variation ($P(r)\sim\frac{2\pi G\rho^2}{3}(R_p^2-r^2)$). On the other hand, the melting range covers $\sim300$\,K (1661\,K--1982\,K at 0 pressure), which contributes to the main uncertainty in $T_{crit}$. Furthermore, \citealp{2015aste.book..533S} suggests a slightly different melting range of $\sim$ 1400\,K--1800\,K, and \citealp{1999M&PS...34..735M} concludes that $T_{solidus}\sim1600$\,K with $T_{crit}\sim1700$\,K based on experiments. Based on these studies, $T_{\mathit{crit}}$ may vary from 1400\,K to 2000\,K by maximum, corresponding to around $10\%$ ($5\%$) difference in $f_{\mathit{Mm}}$ ($f_{\mathit{Mp}}$) as is shown in Figure \ref{mass fraction of diff plot}.

\subsubsection{Thermal penetration}
\label{thermal discussion 2}

Equally important to the critical temperature is the thermal diffusivity $\alpha_d$, which has a strong influence on the ability of heat to penetrate into the planetesimals ($d_{\mathit{diff}}\propto\sqrt{\alpha_d}$), and which we approximate as a constant. In reality, $\alpha$ is a potentially strong function of both the composition of the planetesimals and local temperature. As planetesimals heat up, $\alpha$ likely decreases and thus, could decrease the ability of heat to penetrate during the AGB/RGB. A quick calculation is presented based on the model in \citealp{Miao2014}:

\begin{equation}
\alpha_d=a+\frac{b}{T},
\end{equation}

\noindent where $\alpha_d$ has the unit of $\mathrm{mm^2/s}$, $a$ ranges from $\sim0.13$--0.20 and $b$ ranges from $\sim210$--330 for different types of rocks. There is a factor of $\sim5$ decrease in $\alpha_d$ from our initial condition, $T=150$\,K, to $T_{\mathit{solidus}}\sim 1400$\,K, equivalent to a factor of 2 decrease in the penetration depth during AGB/RGB. This means that the radius of the largest planetesimal that undergoes large-scale melting in planetesimal may be reduced by maximum to 15\,km from 30\,km. Meanwhile, a planetesimal's instantaneous surface temperature is a function of latitude instead of a constant, resulting in asymmetric heating and a loss of spherical symmetry. This temperature contrast can lead to 3-D convection in melted regions and possibly faster thermal penetration. In reality, a thermal evolution model involving not only conduction, but also convection, phase transition, and variations in the local thermal and physical properties may be required to model the evolving interior of the planetesimal, at the cost of longer computational time and additional free parameters, for instance, initial composition of the body \citep{2016ApJ...832..160M}.

\subsubsection{Initial distributions of planetesimals accreted by WDs}
\label{planetesimal distribution}

\begin{figure}
\begin{center}
\includegraphics[width=0.45\textwidth]{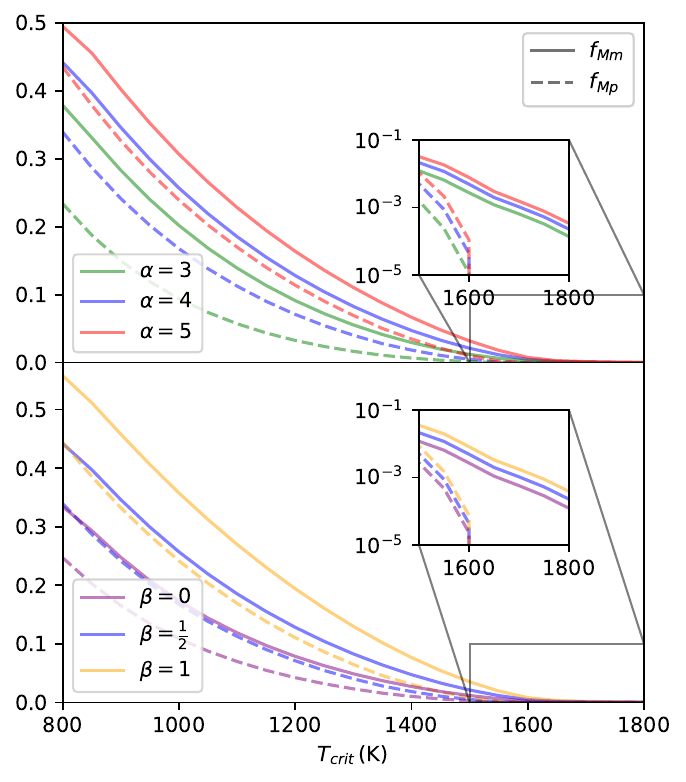}   
\caption{Mass fraction of differentiated samples under different planetesimal size ($\alpha$) and spatial distributions ($\beta$) relative to $\alpha=4$, $\beta=\frac{1}{2}$.}
\label{alpha-beta plot}
\end{center}
\end{figure}

\begin{figure}
\begin{center}
\includegraphics[width=0.45\textwidth]{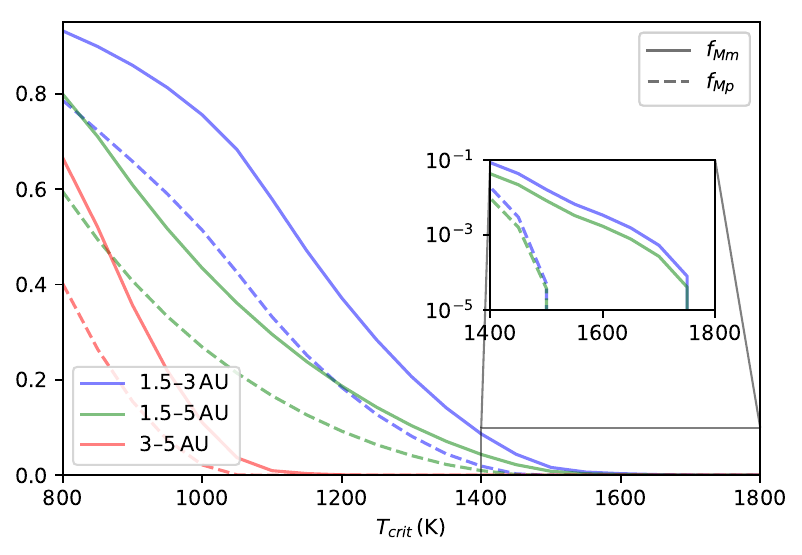}   
\caption{Mass fraction of differentiated samples under different integration range in a 1 solar mass star system ($\alpha=4$, $\beta=\frac{1}{2}$).}
\label{integration range plot}
\end{center}
\end{figure}

Whilst the limits on the maximum size of a planetesimal where melting occurs as a function of distance to the star is robust (given the above assumptions), the fraction of planetesimals accreted by white dwarfs that are core-mantle differentiated is much more uncertain, due to planetesimal distributions and the position/size of the accretion zone. 

We apply $\alpha=4$, $\beta=\frac{1}{2}$ ($\frac{\partial^2 M_{\mathit{total}}}{\partial R_p \partial a_0}\propto R_p^{-1}a_0^{-\frac{1}{2}}$) in the simulations, assuming $R_p$ and $a_0$ are independent variables and the distributions are unaltered by scattering. In terms of spatial distribution ($\beta$), we utilize the result of Minimum Mass Solar Nebula model (MMSN, \citealp{1981PThPS..70...35H,2009ApJ...698..606C}), not necessarily applicable to other systems. In terms of size distribution, $\alpha$, we apply the assumption of collisional evolution \citep{1969JGR....74.2531D,2012ApJ...754...74G,2012ApJ...747..113P}, which is steeper than that of planetesimals formed in the streaming instability \citep{2016ApJ...822...55S,2017ApJ...847L..12S}, with $\alpha\lesssim 3$. There is no guarantee that these values apply to all planetary systems and to all size ranges. Furthermore, the planetary system can alter the distributions via resonances. In Figure \ref{alpha-beta plot}, we show how $\alpha$ and $\beta$ may change our results, $f_{\mathit{Mm}}(T_{\mathit{crit}})$ and $f_{\mathit{Mp}}(T_{\mathit{crit}})$, in a system with a 1 solar mass host star. With the increase in $\alpha$ ($\beta$), more masses of the planetesimal population concentrate towards the host star (are occupied by small-sized bodies), leading to a higher degree of differentiation of the whole planetesimal population. 
 
Scattering and accretion of planetesimals towards the WD rely on the interactions with the planetary system, which can be non-uniform in space and size. For instance, \citealp{2022ApJ...924...61L} simulates the accretion of solar system planetesimals onto the Sun when it becomes a white dwarf and concludes that the geometry of the solar system leads to inside-out accretion (the inner bodies are accreted earlier than their outer counterparts). In Figure \ref{integration range plot}, we present $f_{\mathit{Mm}}(T_{\mathit{crit}})$ and $f_{\mathit{Mp}}(T_{\mathit{crit}})$ for three subsets of planetesimal population described in Section \ref{diff fraction}, initially lying between a: 1.5--3\,AU, b: 1.5--5\,AU, and c: 3--5\,AU around a 1 solar mass host star. In general, $f_{\mathit{Mm}}$ and $f_{\mathit{Mp}}$ decrease rapidly with the outward migration or expansion of the scattering zone. 

In this study, we only consider spherically symmetric planetesimals and neglect their shape distribution, as well as the possible alterations to their shapes after melting. In reality, the shape of a planetesimal may affect its orbital and thermal evolution, as well as the scattering and disruption processes, for instance, disruption outside the Roche limit \citep{2018MNRAS.478L..95K,2020MNRAS.492.5291V,2021MNRAS.506.4031M}, which are beyond the scope of this study.

In the solar system, it is common for asteroids to reside in an elliptical orbit \citep{2017MNRAS.465.4381M}. In this case, Equation \ref{equilibrium temperature equation} only represents instantaneous equilibrium temperature which varies with the true anomaly. We introduce the time-averaged equilibrium temperature ($<T_{\mathit{eq}}>$) for one orbital period ($\tau$) \citep{M_ndez_2017}:

\begin{equation}
\label{time averaged equilibrium temperature equation}
\begin{aligned}
&<T_{\mathit{eq}}>=\frac{1}{\tau}\int_{0}^{\tau} T_{\mathit{eq}}(r)dt\\&=\frac{1}{2\pi}\int_{0}^{2\pi}\left[\frac{L_*(1-A_B)}{16\pi\epsilon\sigma \beta_r a^2(1-e\cos E)^2}\right]^{\frac{1}{4}}(1-e\cos E)dE\\&=T_{\mathit{eq}}(r=a)\frac{2\sqrt{1+e}}{\pi}E_2\left(\sqrt{\frac{2e}{1+e}}\right),
\end{aligned}
\end{equation}

\noindent where  $r=a(1-e\cos E)$, with $E$ the eccentric anomaly, $E_2$ is the complete elliptic integral of the second kind, $T_{\mathit{eq}}(r)=\left(\frac{L_*(1-A_B)}{16\pi\epsilon\sigma \beta_r r^2}\right)^{\frac{1}{4}}$ (Equation \ref{equilibrium temperature equation}) . Compared to $T_{\mathit{eq}}$, $<T_{\mathit{eq}}>$ decreases by $10\%$ for identical semi-major axis and $e\gtrsim$ 0.1. Meanwhile, as the pericentre distance of elliptical orbits ($r_{\mathit{peri}}$) satisfies $r_{\mathit{peri}}=a(1-e)<a$, critical $a_0$ increases compared to that for circular orbits. For identical initial semi-major axis distribution, a population of planetesimals on circular orbits are of the highest degree of differentiation with other conditions kept constant.

\subsubsection{Observability of planetesimals differentiated under stellar irradiation}
\label{observational limit}

\begin{figure}
\includegraphics[width=0.45\textwidth]{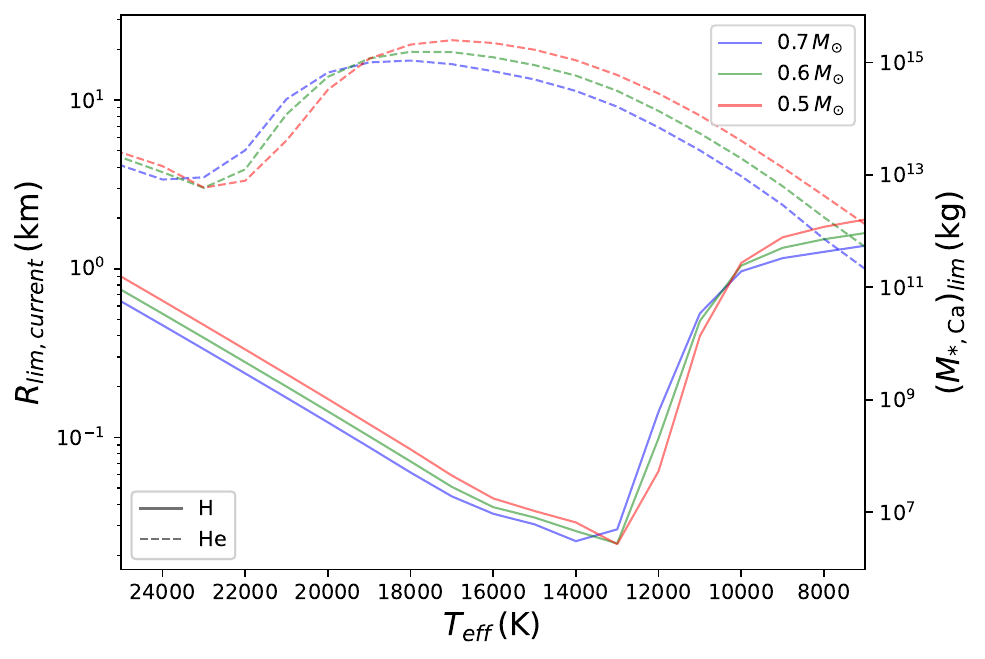}   
\caption{The minimum size of the planetesimal ($R_{\mathit{lim},\mathit{current}}$) and the corresponding mass of calcium in the atmospheres of the white dwarf ($(M_{*,\rm Ca})_{\mathit{lim}}$), that leads to observable photospheric calcium. The mass fraction of calcium is assumed to be $1.5\%$ in the sample planetesimals. We fit a linear relationship between white dwarf's effective temperature ($T_{\mathit{eff}}$) and the detection limit of $\log_{10}\left[\frac{n(\mathrm{Ca)}}{n(\mathrm{Hx})}\right]$, to the transformed Ca/H and Ca/He (original data in \citealp{2003ApJ...596..477Z,2010ApJ...722..725Z}) at a limiting equivalent width of 14 m\r{A} for a S/N of 30 and spectrograph resolution of 40000 at given $T_{\mathit{eff}}$. The kink at $T_{\mathit{eff}}\sim 12000$\,K and 22000\,K for white dwarfs with H and He--dominated atmospheres marks the transition from radiative to convective behaviour \citep{2011ApJ...737...28B,2023MNRAS.519.4529C}.}
\label{limiting size of instantaneously acrreted body plot}
\end{figure}

\begin{figure}
\includegraphics[width=0.45\textwidth]{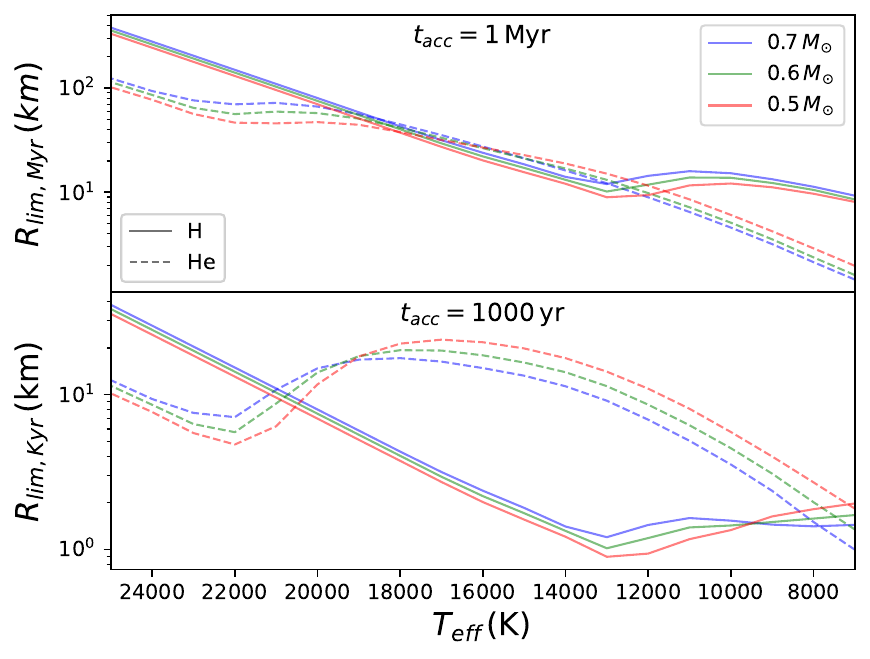}   
\caption{The minimum size of the planetesimal resulting in detected photospheric calcium assuming the same detection limits as Figure \ref{limiting size of instantaneously acrreted body plot} and that the observed metal pollution originates from a single body that accreted onto the white dwarf over 1\,Myr (upper panel) and 1\,Kyr (lower panel) assuming $t=t_{\mathit{acc}}$ in Equation \ref{mass of calcium in the atmosphere equation}, respectively. The mass fraction of calcium is assumed to be $1.5\%$ in the sample planetesimals.}
\label{limiting size plot 1 Myr}
\end{figure}

The smallest planetesimals, which the model predicts to undergo the largest degree of melting, also produce the weakest signal when accreted, and are not necessarily observable throughout the cooling age of the white dwarf. Our ability to detect calcium (Ca) in a white dwarf atmosphere depends crucially on the signal to noise (S/N) achieved in the observed spectra, alongside the spectral resolution. In order to investigate whether the planetesimals differentiated under stellar irradiation are detectable, we compute the minimum detectable atmospheric Ca/H and Ca/He of the polluted white dwarfs in \citealp{2003ApJ...596..477Z,2010ApJ...722..725Z} at an equivalent width of 14\,m\r{A} (the limiting equivalent width required to resolve the Ca lines for S/N of 30 and spectrograph resolution of 40000), and find the linear relation between the minimum detectable relative abundances of calcium ($\log_{10}\left[\frac{n(\mathrm{Ca)}}{n(\mathrm{Hx})}\right]_{\mathit{lim}}$) and the white dwarf's effective temperature ($T_{\mathit{eff}}$):

\begin{equation}\label{linear relation}
 \log_{10}\left[\frac{n(\mathrm{Ca)}}{n(\mathrm{Hx})}\right]_{lim}=m\frac{T_{\mathit{eff}}}{K}+b,   
\end{equation}

\noindent where $n(\mathrm{Ca})$ is the number density of calcium and $n(\rm Hx)$ is the number density of hydrogen/helium in the atmosphere of the white dwarf. $m\approx 3.8\times10^{-4} (4.2\times10^{-4})$, $b\approx -14.2 (-16.6)$ for hydrogen (helium) atmospheres. The minimum mass of calcium in the atmosphere of the white dwarf ($(M_{*,\rm Ca})_{\mathit{lim}}$) that is observable is of the form:

\begin{equation}\label{instantaneous obs mass equation}
(M_{*,\rm Ca})_{\mathit{lim}}=\left[\frac{n(\mathrm{Ca})}{n(\mathrm{Hx})}\right]_{\mathit{lim}}\frac{A_{\mathrm{Ca}}}{A_{\mathrm{Hx}}}M_{*}q, 
\end{equation}

\noindent where $q$ is the ratio of white dwarf's atmosphere mass to the white dwarf mass, obtained from Montreal White Dwarf Database (MWDD, \citealp{2017ASPC..509....3D,2020ApJ...901...93B}), $A_{\rm Ca}$ and $A_{\rm Hx}$ are atomic masses of calcium and hydrogen/helium, respectively. We can estimate the minimum size of the individually accreted body ($R_{\mathit{lim}}$) that can explain the current atmospheric Ca abundance of the white dwarf by:

\begin{equation}\label{instantaneous obs limit equation}
R_{\mathit{lim}}=\left(\frac{M_{*,\rm Ca}}{4\pi \rho f_{m,\rm Ca}}\right)^{\frac{1}{3}},
\end{equation}

\noindent where $M_{*,\rm Ca}$ is the mass of Ca in the atmosphere of the white dwarf. We assume that sample planetesimals have a bulk-Earth like calcium mass fraction $f_{m,\rm Ca}\approx1.5\%$.

The observational limit (minimum size of the accreted body $R_{\mathit{lim},\mathit{current}}$, and minimum calcium mass in the atmosphere of the white dwarf, $(M_{*,\rm Ca})_{\mathit{lim}}$ that can lead to observable signals) as a function of WD's effective temperature calculated from Equation \ref{linear relation}, \ref{instantaneous obs mass equation} and \ref{instantaneous obs limit equation} is shown in Figure \ref{limiting size of instantaneously acrreted body plot} for WDs with $M_*=0.5$, 0.6, 0.7\,$M_{\odot}$ and H, He-dominated atmospheres, respectively. Observations of instantaneous accretion are more sensitive to WDs with H-dominated atmosphere until $T_{\mathit{eff}}\lesssim8000$\,K ($R_{\mathit{lim}}\sim1$\,km), after which $R_{\mathit{lim}}$ of WDs with He-dominated atmosphere is lower. 

In reality, planetesimals are unlikely to accrete instantaneously, rather spreading their accretion over a longer time period that we call $t_{\mathit{acc}}$ (we define the start of accretion to be $t=0$, such that $t_{\mathit{acc}}$ also marks the time when the accretion stops). This is important to consider, as the abundance of Ca in the atmosphere of the white dwarf ($M_{*,\rm Ca}$) has a different relationship to the accretion rate ($\dot{M}_p$, assumed to be constant) before and after $t=t_{\mathit{acc}}$ \citep{2009A&A...498..517K,2014AREPS..42...45J}:

\begin{equation}\label{mass of calcium in the atmosphere equation}
M_{*,\rm Ca}(t)= 
\begin{cases}
\dot{M}_p{f_{m,\rm Ca}t_{\rm Ca}\left(1-e^{-\frac{t}{t_{\rm Ca}}}\right)} & t\leq t_{\mathit{acc}}  \\
\dot{M}_p{f_{m,\rm Ca}t_{\rm Ca}\left(1-e^{-\frac{t_{\mathit{acc}}}{t_{\rm Ca}}}\right)}e^{-\frac{t-t_{\mathit{acc}}}{t_{\rm Ca}}} & t>t_{\mathit{acc}}
\end{cases}
,
\end{equation}

\noindent where $t_{\rm Ca}$ is the sinking timescale of Ca obtained from MWDD \citep{1986ApJS...61..177P}. We consider a special case where $t=t_{\mathit{acc}}$ (the maximum of $M_{*,\rm Ca}(t)$) in this study and estimate the minimum size of a planetesimal ($R_{\mathit{lim}}$) that produces observable Ca lines: 

\begin{equation}
\begin{aligned}
R_{\mathit{lim}}=\left(\frac{3\dot{M}_{p,\mathit{lim}}t_{\mathit{acc}}}{4\pi \rho }\right)^{\frac{1}{3}},
\end{aligned}
\end{equation}

\noindent where $\dot{M}_{p,\mathit{lim}}=\frac{(M_{*,\rm Ca})_{\mathit{lim}}}{\left(1-e^{-\frac{t_{\mathit{acc}}}{t_{\rm Ca}}}\right)}$.

By including the sinking timescale of Ca and the accretion timescale of planetesimal debris, we show the corresponding observational limit ($R_{\mathit{lim}}$) in Figure \ref{limiting size plot 1 Myr} with 2 distinct $t_{\mathit{acc}}$, i: 1\,Myr (upper panel) and ii: 1\,Kyr (lower panel) in the same parameter space as Figure \ref{limiting size of instantaneously acrreted body plot}. 

In white dwarfs where the pollutants accrete over 1\,Myr, the minimum size of a planetesimal that leads to observable Ca lines, $R_{lim}$ decreases by $\approx$ 2 orders of magnitude from $T_{\mathit{eff}}=25000$\,K to 7000\,K for white dwarfs with both H and He-dominated atmospheres. If the pollutants come from a single body, accretion of a body with $R_p\lesssim 10$\,km is only detectable in a white dwarf with He-dominated atmosphere and $T_{eff} \lesssim10000$\,K.

If, instead, pollutants accrete over 1000 yr, the predicted $R_{lim}$ approaches that of instantaneous accretion (Figure \ref{limiting size of instantaneously acrreted body plot}) at $T_{\mathit{eff}}\lesssim7000$\,K (20000\,K) for white dwarfs with a H (He)-dominated atmospheres. These transitions occur at $t_{Ca}\gg t_{acc}$, such that the mass of white dwarf pollutants at the end of accretion is representative of the total mass of the accreted body. The dependence of $R_{\mathit{lim}}$ on $t_{\mathit{acc}}$ is stronger for white dwarfs with H-dominated atmosphere because of their much shorter $t_{\rm Ca}$ compared to their counterparts with He-dominated atmosphere at identical $T_{\mathit{eff}}$ (about 2--3 orders of magnitude).

Large-scale melting ($\gtrsim 95\%$ of the volume) triggered by stellar evolution, as is shown in Figure \ref{volume fraction of diff plot}, can only occur in planetesimals with $R_p\lesssim30$\,km even when $T_{\mathit{crit}}$ is lowered to 1400\,K. Figure \ref{limiting size of instantaneously acrreted body plot} and \ref{limiting size plot 1 Myr} illustrate that those small ($R_p\lesssim 30$\,km) planetesimals potentially undergo large-scale melting induced by giant branch evolution are only detectable (assuming single body accretion) in the scenario that the body accretes rapidly ($t_{\mathit{acc}}\lesssim1$\,Kyr) or the white dwarf cools down to $T_{eff}$ $\lesssim$ 16000 K. 

For He-white dwarfs cooler than $\sim16000$\,K, it is possible to detect planetesimals as small as $\sim 30$\,km, with this limit decreasing to $\sim1$\,km when the white dwarf cools to $T_{eff}=7000$\,K. For these white dwarfs, more smaller-sized bodies, potentially being more thermally processed become observable, when accreted individually. However, the coolest He-white dwarfs also have extremely long Ca sinking timescale, which adds the possibility that the pollutants consist of several accreted bodies. 

Compared to the preferred parameter space where planetesimals differentiate under radiogenic heating ($R_p\gtrsim50$\,km  \citep{2022MNRAS.515..395C}), stellar irradiation preferentially differentiate smaller-sized bodies. Therefore, irradiation-differentiated bodies become observable later in the cooling age of the white dwarf, with a definitive upper bound of the absolute pollutant abundances equal to the largest body that may differentiate under stellar irradiation (30 km according our model).

\subsection{Implications}
\label{implications}

\subsubsection{Do we observe post-main sequence differentiation?}
\label{inferred fraction of differentiation}

Our results show that over-abundances of siderophile/lithophile elements in the atmosphere of a cool white dwarf can result from a parent body differentiated by stellar irradiation, but only in the case that the absolute abundances of WD pollutants drops below that of a 30\,km planetesimal. 

In this section we further discuss if any observed white dwarfs, whose pollutants are identified as core/mantle-rich, could result from accretion of small-sized planetesimals differentiated during stellar evolution. In order to investigate this we make the broad assumption that the observed atmospheric abundances are dominated by a single body, noting that this may not be the case \citep{10.1093/mnras/stz3191}. 

The majority of white dwarfs with helium-dominated atmospheres identified in the literature as accreting core or mantle-rich (or even crust-rich) material are highly polluted (e.g., SDSSJ0736+4118, SDSSJ0744+4649, \citealp{2017MNRAS.467.4970H,2021MNRAS.504.2853H}). Such white dwarfs have likely accreted a body of $\gtrsim$ 50 km in radius to match the observed Ca abundances. Optimistically, a 50\,km planetesimal can melt $\sim 40\%$ of its volume under stellar irradiation (Figure \ref{volume fraction of diff plot}, right panels). This low degree of differentiation is unlikely to explain the visible siderophile/lithophile character of these white dwarf pollutants. In this case, radiogenic heating, which preferentially differentiates planetesimals with $R_p\gtrsim$ 50\,km \citep{2022MNRAS.515..395C} may be a better explanation. A notable exception among these He-white dwarfs with core/mantle-rich pollutants is WD0122-227 whose abundances were interpreted as core-rich by \citealp{2019MNRAS.490..202S,2022MNRAS.510.3512B}. The $\log_{10}\left[\frac{n(\mathrm{Ca)}}{n(\mathrm{He})}\right]$ of -10.1 corresponds to the accretion of a $\sim30$\,km planetesimal, which if it resided interior to $\sim1.5$\,AU of a 1--3\,$M_{\odot}$ star and melted at 1400\,K, would have undergone large-scale melting on the giant branches.

Among the white dwarfs with H-dominated atmospheres, WD2105-820, PG0843+516 \citealp{2019MNRAS.490..202S,2020ApJ...898...84K}  have been identified as core/mantle-rich and have low metal abundances in their atmospheres, equivalent to accretion of a $\lesssim1$\,km planetesimal. These weakly polluted systems are consistent with the instantaneous accretion of planetesimals sufficiently small to have been differentiated by giant branch heating. However, the uncertainty lies in the short sinking timescale of Ca (ranging from $\sim 10^{-2}$\,yr to 10\,yr) in the H-dominated atmospheres of white dwarfs, adding the probability that current Ca mass is well below that contained in the parent body whose accretion continues over many sinking timescales.

In general, we do not expect the core/mantle-rich pollutants of white dwarfs with He-dominated atmospheres presented in the literature to date to result from planetesimals differentiated during giant branches. Considering the much longer sinking timescale compared to H-white dwarfs, future observations targeting at pollutants in the He-dominated atmospheres of white dwarfs with over-abundances in siderophile/lithophile elements but low absolute abundances of Ca may provide more evidence for stellar-induced differentiation.

\subsubsection{Depletion of moderately volatiles }
\label{volatile loss and retention}

The degree of depletion in moderately volatiles, including elements such as Mn, Na, is considered as a powerful tool to probe the early thermal history of a planetesimal \citep{2014AREPS..42...45J}. For instance, bulk Earth is depleted in these species relative to chondritic meteorites \citep{2018MNRAS.479.3814H}, most likely explained by the incomplete condensation of Na-bearing minerals out of the nebula gas early in planet formation \citep{2008RSPTA.366.4205O,2014PNAS..11117029P,SIEBERT2018130}. Planetary materials accreted by white dwarfs are often Na-poor compared to solar compositions. \citep{2020ApJ...901...10D,2021MNRAS.504.2853H} and \citealp{2021E&PSL.55416694H} argue that depletion in moderately volatiles could occur early in planet formation, due to incomplete condensation of the nebula gas, or after the nebula gas has dissipated due to large-scale melting in magma oceans following impacts or radiogenic heating. 

For those planetesimals that undergo large-scale melting to form a magma ocean induced by stellar irradiation on the giant branches (Section \ref{volume fraction of melting}), additional loss of moderately volatiles such as Na and Mn is anticipated, as well as segregation of the iron from the silicates. For those planetesimals heated insufficiently for the silicates to melt and iron to mobilise, depletion of moderately volatiles such as Na largely depends on the thermal properties of the host minerals. Hence, it is unclear whether and how much the moderately volatiles will be lost. If Na is hosted in refractory silicates, whose melting is necessary for Na to be movable, the critical temperature of Na depletion may be coincide with/exceed the melting temperature of silicates. In this case, the observed white dwarf pollutants depleted in moderately volatiles is a signature of heating during the planetary formation stage \citep{2021E&PSL.55416694H}. Meanwhile, if Na (partially) exists in the minerals of lower thermal stability, as proposed by \citealp{2021arXiv210807331M} who suggest Na hosted in sodalite or nepheline could degas above $\sim1000$\,K, although a high level of moderately volatile-element depletion in massive white dwarf pollutants almost certainly originate from heating around planetary formation, it is unclear whether stellar irradiation could be the cause of moderately volatile-element depletion in the less massive ($\lesssim$ mass of a 50\,km planetesimal) white dwarf pollutants.

\section{Conclusion}
\label{conclusion}

Many white dwarfs have accreted planetary materials that are rich in either core or mantle material. The formation of iron cores requires a period of large-scale melting. This work shows that very few planetesimals orbiting white dwarfs underwent large-scale melting triggered by stellar irradiation on the giant branches. For a solar like host star, a planetesimal must initially reside between $\sim1.3$\,AU and 1.7\,AU and be smaller than $\sim30$\,km in radius in order to be melted over $95\%$ of the volume and survive the asymptotic giant branch of the host star, even when the critical temperature of melting is lowered to 1400\,K.

This work highlights planetesimal size as a key differentiator between large-scale melting that occurs due to heating on the giant branches and that due to the decay of ${}^{26}\mathrm{Al}$ soon after planet formation. For Solar System abundances of ${}^{26}\mathrm{Al}$, the latter prefers bodies $\gtrsim50$\,km in radius, whilst the former occurs only in bodies where $R_p\lesssim30$\,km. Thus, accretion of these small planetesimals that underwent large-scale melting due to giant branch heating are most likely to be seen in cool white dwarfs, especially those with relatively low absolute abundances of pollutants. They do not represent the current observed population of core/mantle-rich white dwarf pollutants.

\section*{Acknowledgements}

We would like to thank Andrew Buchan and Laura Rogers for providing relation \ref{linear relation} and summarizing the white dwarf data used in Section \ref{implications}. AB acknowledges the support of a Royal Society University Research Fellowship, URF\textbackslash R1\textbackslash 211421. YL acknowledges the support of a STFC studentship. 

\section*{Data Availability}

Stellar equivalent evolutionary tracks from MIST is available at \url{https://waps.cfa.harvard.edu/MIST/}. MESA stellar evolution code is available at \url{https://docs.mesastar.org/en/release-r23.05.1/}. Other codes and data used in this work are available upon reasonable request to the author, Yuqi Li. 



\bibliographystyle{mnras}
\bibliography{example} 

\begin{thebibliography}{}
\makeatletter
\relax
\def\mn@urlcharsother{\let\do\@makeother \do\$\do\&\do\#\do\^\do\_\do\%\do\~}
\def\mn@doi{\begingroup\mn@urlcharsother \@ifnextchar [ {\mn@doi@} {\mn@doi@[]}}
\def\mn@doi@[#1]#2{\def\@tempa{#1}\ifx\@tempa\@empty \href {http://dx.doi.org/#2} {doi:#2}\else \href {http://dx.doi.org/#2} {#1}\fi \endgroup}
\def\mn@eprint#1#2{\mn@eprint@#1:#2::\@nil}
\def\mn@eprint@arXiv#1{\href {http://arxiv.org/abs/#1} {{\tt arXiv:#1}}}
\def\mn@eprint@dblp#1{\href {http://dblp.uni-trier.de/rec/bibtex/#1.xml} {dblp:#1}}
\def\mn@eprint@#1:#2:#3:#4\@nil{\def\@tempa {#1}\def\@tempb {#2}\def\@tempc {#3}\ifx \@tempc \@empty \let \@tempc \@tempb \let \@tempb \@tempa \fi \ifx \@tempb \@empty \def\@tempb {arXiv}\fi \@ifundefined {mn@eprint@\@tempb}{\@tempb:\@tempc}{\expandafter \expandafter \csname mn@eprint@\@tempb\endcsname \expandafter{\@tempc}}}

\bibitem[\protect\citeauthoryear{{B{\'e}dard}, {Bergeron}, {Brassard}  \& {Fontaine}}{{B{\'e}dard} et~al.}{2020}]{2020ApJ...901...93B}
{B{\'e}dard} A.,  {Bergeron} P.,  {Brassard} P.,   {Fontaine} G.,  2020, \mn@doi [\apj] {10.3847/1538-4357/abafbe}, \href {https://ui.adsabs.harvard.edu/abs/2020ApJ...901...93B} {901, 93}

\bibitem[\protect\citeauthoryear{{Bergeron} et~al.,}{{Bergeron} et~al.}{2011}]{2011ApJ...737...28B}
{Bergeron} P.,  et~al., 2011, \mn@doi [\apj] {10.1088/0004-637X/737/1/28}, \href {https://ui.adsabs.harvard.edu/abs/2011ApJ...737...28B} {737, 28}

\bibitem[\protect\citeauthoryear{{Bochkarev} \& {Rafikov}}{{Bochkarev} \& {Rafikov}}{2011}]{2011ApJ...741...36B}
{Bochkarev} K.~V.,  {Rafikov} R.~R.,  2011, \mn@doi [\apj] {10.1088/0004-637X/741/1/36}, \href {https://ui.adsabs.harvard.edu/abs/2011ApJ...741...36B} {741, 36}

\bibitem[\protect\citeauthoryear{{Bonsor}, {Carter}, {Hollands}, {G{\"a}nsicke}, {Leinhardt}  \& {Harrison}}{{Bonsor} et~al.}{2020}]{2020MNRAS.492.2683B}
{Bonsor} A.,  {Carter} P.~J.,  {Hollands} M.,  {G{\"a}nsicke} B.~T.,  {Leinhardt} Z.,   {Harrison} J. H.~D.,  2020, \mn@doi [\mnras] {10.1093/mnras/stz3603}, \href {https://ui.adsabs.harvard.edu/abs/2020MNRAS.492.2683B} {492, 2683}

\bibitem[\protect\citeauthoryear{{Brouwers}, {Bonsor}  \& {Malamud}}{{Brouwers} et~al.}{2023}]{2023MNRAS.519.2646B}
{Brouwers} M.~G.,  {Bonsor} A.,   {Malamud} U.,  2023, \mn@doi [\mnras] {10.1093/mnras/stac3316}, \href {https://ui.adsabs.harvard.edu/abs/2023MNRAS.519.2646B} {519, 2646}

\bibitem[\protect\citeauthoryear{{Buchan}, {Bonsor}, {Shorttle}, {Wade}, {Harrison}, {Noack}  \& {Koester}}{{Buchan} et~al.}{2022}]{2022MNRAS.510.3512B}
{Buchan} A.~M.,  {Bonsor} A.,  {Shorttle} O.,  {Wade} J.,  {Harrison} J.,  {Noack} L.,   {Koester} D.,  2022, \mn@doi [\mnras] {10.1093/mnras/stab3624}, \href {https://ui.adsabs.harvard.edu/abs/2022MNRAS.510.3512B} {510, 3512}

\bibitem[\protect\citeauthoryear{{Caricchi}, {Burlini}, {Ulmer}, {Gerya}, {Vassalli}  \& {Papale}}{{Caricchi} et~al.}{2007}]{2007E&PSL.264..402C}
{Caricchi} L.,  {Burlini} L.,  {Ulmer} P.,  {Gerya} T.,  {Vassalli} M.,   {Papale} P.,  2007, \mn@doi [Earth and Planetary Science Letters] {10.1016/j.epsl.2007.09.032}, \href {https://ui.adsabs.harvard.edu/abs/2007E&PSL.264..402C} {264, 402}

\bibitem[\protect\citeauthoryear{{Caron}, {Bergeron}, {Blouin}  \& {Leggett}}{{Caron} et~al.}{2023}]{2023MNRAS.519.4529C}
{Caron} A.,  {Bergeron} P.,  {Blouin} S.,   {Leggett} S.~K.,  2023, \mn@doi [\mnras] {10.1093/mnras/stac3733}, \href {https://ui.adsabs.harvard.edu/abs/2023MNRAS.519.4529C} {519, 4529}

\bibitem[\protect\citeauthoryear{{Chambers}}{{Chambers}}{2004}]{2004E&PSL.223..241C}
{Chambers} J.~E.,  2004, \mn@doi [Earth and Planetary Science Letters] {10.1016/j.epsl.2004.04.031}, \href {https://ui.adsabs.harvard.edu/abs/2004E&PSL.223..241C} {223, 241}

\bibitem[\protect\citeauthoryear{{Choi}, {Dotter}, {Conroy}, {Cantiello}, {Paxton}  \& {Johnson}}{{Choi} et~al.}{2016}]{2016ApJ...823..102C}
{Choi} J.,  {Dotter} A.,  {Conroy} C.,  {Cantiello} M.,  {Paxton} B.,   {Johnson} B.~D.,  2016, \mn@doi [\apj] {10.3847/0004-637X/823/2/102}, \href {https://ui.adsabs.harvard.edu/abs/2016ApJ...823..102C} {823, 102}

\bibitem[\protect\citeauthoryear{{Costa}}{{Costa}}{2005}]{2005GeoRL..3222308C}
{Costa} A.,  2005, \mn@doi [\grl] {10.1029/2005GL024303}, \href {https://ui.adsabs.harvard.edu/abs/2005GeoRL..3222308C} {32, L22308}

\bibitem[\protect\citeauthoryear{{Crida}}{{Crida}}{2009}]{2009ApJ...698..606C}
{Crida} A.,  2009, \mn@doi [\apj] {10.1088/0004-637X/698/1/606}, \href {https://ui.adsabs.harvard.edu/abs/2009ApJ...698..606C} {698, 606}

\bibitem[\protect\citeauthoryear{{Cristallo}, {Straniero}, {Piersanti}  \& {Gobrecht}}{{Cristallo} et~al.}{2015}]{2015ApJS..219...40C}
{Cristallo} S.,  {Straniero} O.,  {Piersanti} L.,   {Gobrecht} D.,  2015, \mn@doi [\apjs] {10.1088/0067-0049/219/2/40}, \href {https://ui.adsabs.harvard.edu/abs/2015ApJS..219...40C} {219, 40}

\bibitem[\protect\citeauthoryear{{Curry}, {Bonsor}, {Lichtenberg}  \& {Shorttle}}{{Curry} et~al.}{2022}]{2022MNRAS.515..395C}
{Curry} A.,  {Bonsor} A.,  {Lichtenberg} T.,   {Shorttle} O.,  2022, \mn@doi [\mnras] {10.1093/mnras/stac1709}, \href {https://ui.adsabs.harvard.edu/abs/2022MNRAS.515..395C} {515, 395}

\bibitem[\protect\citeauthoryear{{Dohnanyi}}{{Dohnanyi}}{1969}]{1969JGR....74.2531D}
{Dohnanyi} J.~S.,  1969, \mn@doi [\jgr] {10.1029/JB074i010p02531}, \href {https://ui.adsabs.harvard.edu/abs/1969JGR....74.2531D} {74, 2531}

\bibitem[\protect\citeauthoryear{{Dorn}, {Khan}, {Heng}, {Alibert}, {Connolly}, {Benz}  \& {Tackley}}{{Dorn} et~al.}{2015}]{2015arXiv150203605D}
{Dorn} C.,  {Khan} A.,  {Heng} K.,  {Alibert} Y.,  {Connolly} J. A.~D.,  {Benz} W.,   {Tackley} P.,  2015, \mn@doi [arXiv e-prints] {10.48550/arXiv.1502.03605}, \href {https://ui.adsabs.harvard.edu/abs/2015arXiv150203605D} {p. arXiv:1502.03605}

\bibitem[\protect\citeauthoryear{{Dotter}}{{Dotter}}{2016}]{2016ApJS..222....8D}
{Dotter} A.,  2016, \mn@doi [\apjs] {10.3847/0067-0049/222/1/8}, \href {https://ui.adsabs.harvard.edu/abs/2016ApJS..222....8D} {222, 8}

\bibitem[\protect\citeauthoryear{{Doyle}, {Young}, {Klein}, {Zuckerman}  \& {Schlichting}}{{Doyle} et~al.}{2019}]{2019Sci...366..356D}
{Doyle} A.~E.,  {Young} E.~D.,  {Klein} B.,  {Zuckerman} B.,   {Schlichting} H.~E.,  2019, \mn@doi [Science] {10.1126/science.aax3901}, \href {https://ui.adsabs.harvard.edu/abs/2019Sci...366..356D} {366, 356}

\bibitem[\protect\citeauthoryear{{Doyle}, {Klein}, {Schlichting}  \& {Young}}{{Doyle} et~al.}{2020}]{2020ApJ...901...10D}
{Doyle} A.~E.,  {Klein} B.,  {Schlichting} H.~E.,   {Young} E.~D.,  2020, \mn@doi [\apj] {10.3847/1538-4357/abad9a}, \href {https://ui.adsabs.harvard.edu/abs/2020ApJ...901...10D} {901, 10}

\bibitem[\protect\citeauthoryear{{Dufour}, {Blouin}, {Coutu}, {Fortin-Archambault}, {Thibeault}, {Bergeron}  \& {Fontaine}}{{Dufour} et~al.}{2017}]{2017ASPC..509....3D}
{Dufour} P.,  {Blouin} S.,  {Coutu} S.,  {Fortin-Archambault} M.,  {Thibeault} C.,  {Bergeron} P.,   {Fontaine} G.,  2017, in {Tremblay} P.~E.,  {Gaensicke} B.,   {Marsh} T.,  eds,  Astronomical Society of the Pacific Conference Series Vol. 509, 20th European White Dwarf Workshop. p.~3 (\mn@eprint {arXiv} {1610.00986}), \mn@doi{10.48550/arXiv.1610.00986}

\bibitem[\protect\citeauthoryear{Falkovich}{Falkovich}{2018}]{falkovich_2018}
Falkovich G.,  2018, Basic notions and steady flows, 2 edn.
Cambridge University Press, p. 1–62, \mn@doi{10.1017/9781316416600.003}

\bibitem[\protect\citeauthoryear{{Farihi}, {G{\"a}nsicke}  \& {Koester}}{{Farihi} et~al.}{2013}]{2013Sci...342..218F}
{Farihi} J.,  {G{\"a}nsicke} B.~T.,   {Koester} D.,  2013, \mn@doi [Science] {10.1126/science.1239447}, \href {https://ui.adsabs.harvard.edu/abs/2013Sci...342..218F} {342, 218}

\bibitem[\protect\citeauthoryear{{Ferich}, {Baronett}, {Tamayo}  \& {Steffen}}{{Ferich} et~al.}{2022}]{2022ApJS..262...41F}
{Ferich} N.,  {Baronett} S.~A.,  {Tamayo} D.,   {Steffen} J.~H.,  2022, \mn@doi [\apjs] {10.3847/1538-4365/ac8d60}, \href {https://ui.adsabs.harvard.edu/abs/2022ApJS..262...41F} {262, 41}

\bibitem[\protect\citeauthoryear{{G{\'a}sp{\'a}r}, {Psaltis}, {Rieke}  \& {{\"O}zel}}{{G{\'a}sp{\'a}r} et~al.}{2012}]{2012ApJ...754...74G}
{G{\'a}sp{\'a}r} A.,  {Psaltis} D.,  {Rieke} G.~H.,   {{\"O}zel} F.,  2012, \mn@doi [\apj] {10.1088/0004-637X/754/1/74}, \href {https://ui.adsabs.harvard.edu/abs/2012ApJ...754...74G} {754, 74}

\bibitem[\protect\citeauthoryear{Gerya}{Gerya}{2019}]{gerya_2019}
Gerya T.,  2019, Numerical solution of the heat conservation equation, 2 edn.
Cambridge University Press, p. 139–155, \mn@doi{10.1017/9781316534243.011}

\bibitem[\protect\citeauthoryear{{Hansen} \& {Kawaler}}{{Hansen} \& {Kawaler}}{1994}]{1994sipp.book.....H}
{Hansen} C.~J.,  {Kawaler} S.~D.,  1994, {Stellar Interiors. Physical Principles, Structure, and Evolution.}, \mn@doi{10.1007/978-1-4419-9110-2.
}

\bibitem[\protect\citeauthoryear{{Harrison}, {Bonsor}  \& {Madhusudhan}}{{Harrison} et~al.}{2018}]{2018MNRAS.479.3814H}
{Harrison} J. H.~D.,  {Bonsor} A.,   {Madhusudhan} N.,  2018, \mn@doi [\mnras] {10.1093/mnras/sty1700}, \href {https://ui.adsabs.harvard.edu/abs/2018MNRAS.479.3814H} {479, 3814}

\bibitem[\protect\citeauthoryear{{Harrison}, {Bonsor}, {Kama}, {Buchan}, {Blouin}  \& {Koester}}{{Harrison} et~al.}{2021a}]{2021MNRAS.504.2853H}
{Harrison} J. H.~D.,  {Bonsor} A.,  {Kama} M.,  {Buchan} A.~M.,  {Blouin} S.,   {Koester} D.,  2021a, \mn@doi [\mnras] {10.1093/mnras/stab736}, \href {https://ui.adsabs.harvard.edu/abs/2021MNRAS.504.2853H} {504, 2853}

\bibitem[\protect\citeauthoryear{{Harrison}, {Shorttle}  \& {Bonsor}}{{Harrison} et~al.}{2021b}]{2021E&PSL.55416694H}
{Harrison} J. H.~D.,  {Shorttle} O.,   {Bonsor} A.,  2021b, \mn@doi [Earth and Planetary Science Letters] {10.1016/j.epsl.2020.116694}, \href {https://ui.adsabs.harvard.edu/abs/2021E&PSL.55416694H} {554, 116694}

\bibitem[\protect\citeauthoryear{{Hayashi}}{{Hayashi}}{1981}]{1981PThPS..70...35H}
{Hayashi} C.,  1981, \mn@doi [Progress of Theoretical Physics Supplement] {10.1143/PTPS.70.35}, \href {https://ui.adsabs.harvard.edu/abs/1981PThPS..70...35H} {70, 35}

\bibitem[\protect\citeauthoryear{{Hollands}, {Koester}, {Alekseev}, {Herbert}  \& {G{\"a}nsicke}}{{Hollands} et~al.}{2017}]{2017MNRAS.467.4970H}
{Hollands} M.~A.,  {Koester} D.,  {Alekseev} V.,  {Herbert} E.~L.,   {G{\"a}nsicke} B.~T.,  2017, \mn@doi [\mnras] {10.1093/mnras/stx250}, \href {https://ui.adsabs.harvard.edu/abs/2017MNRAS.467.4970H} {467, 4970}

\bibitem[\protect\citeauthoryear{{Hoskin} et~al.,}{{Hoskin} et~al.}{2020}]{2020MNRAS.499..171H}
{Hoskin} M.~J.,  et~al., 2020, \mn@doi [\mnras] {10.1093/mnras/staa2717}, \href {https://ui.adsabs.harvard.edu/abs/2020MNRAS.499..171H} {499, 171}

\bibitem[\protect\citeauthoryear{{Hurley}, {Pols}  \& {Tout}}{{Hurley} et~al.}{2013}]{2013ascl.soft03015H}
{Hurley} J.~R.,  {Pols} O.~R.,   {Tout} C.~A.,  2013, {SSE: Single Star Evolution}, Astrophysics Source Code Library, record ascl:1303.015 (\mn@eprint {ascl} {1303.015})

\bibitem[\protect\citeauthoryear{{Irwin}}{{Irwin}}{2012}]{2012ascl.soft11002I}
{Irwin} A.~W.,  2012, {FreeEOS: Equation of State for stellar interiors calculations}, Astrophysics Source Code Library, record ascl:1211.002 (\mn@eprint {ascl} {1211.002})

\bibitem[\protect\citeauthoryear{{Ivezi{\'c}} et~al.,}{{Ivezi{\'c}} et~al.}{2001}]{2001AJ....122.2749I}
{Ivezi{\'c}} {\v{Z}}.,  et~al., 2001, \mn@doi [\aj] {10.1086/323452}, \href {https://ui.adsabs.harvard.edu/abs/2001AJ....122.2749I} {122, 2749}

\bibitem[\protect\citeauthoryear{{Jermyn}, {Schwab}, {Bauer}, {Timmes}  \& {Potekhin}}{{Jermyn} et~al.}{2021}]{2021ApJ...913...72J}
{Jermyn} A.~S.,  {Schwab} J.,  {Bauer} E.,  {Timmes} F.~X.,   {Potekhin} A.~Y.,  2021, \mn@doi [\apj] {10.3847/1538-4357/abf48e}, \href {https://ui.adsabs.harvard.edu/abs/2021ApJ...913...72J} {913, 72}

\bibitem[\protect\citeauthoryear{{Jermyn} et~al.,}{{Jermyn} et~al.}{2023}]{2023ApJS..265...15J}
{Jermyn} A.~S.,  et~al., 2023, \mn@doi [\apjs] {10.3847/1538-4365/acae8d}, \href {https://ui.adsabs.harvard.edu/abs/2023ApJS..265...15J} {265, 15}

\bibitem[\protect\citeauthoryear{Jia \& Spruit}{Jia \& Spruit}{2018}]{Jia_2018}
Jia S.,  Spruit H.~C.,  2018, \mn@doi [The Astrophysical Journal] {10.3847/1538-4357/aad77c}, 864, 169

\bibitem[\protect\citeauthoryear{{Johansen}, {Ronnet}, {Schiller}, {Deng}  \& {Bizzarro}}{{Johansen} et~al.}{2023}]{2023A&A...671A..74J}
{Johansen} A.,  {Ronnet} T.,  {Schiller} M.,  {Deng} Z.,   {Bizzarro} M.,  2023, \mn@doi [\aap] {10.1051/0004-6361/202142141}, \href {https://ui.adsabs.harvard.edu/abs/2023A&A...671A..74J} {671, A74}

\bibitem[\protect\citeauthoryear{{Jura} \& {Xu}}{{Jura} \& {Xu}}{2010}]{2010AJ....140.1129J}
{Jura} M.,  {Xu} S.,  2010, \mn@doi [\aj] {10.1088/0004-6256/140/5/1129}, \href {https://ui.adsabs.harvard.edu/abs/2010AJ....140.1129J} {140, 1129}

\bibitem[\protect\citeauthoryear{{Jura} \& {Young}}{{Jura} \& {Young}}{2014}]{2014AREPS..42...45J}
{Jura} M.,  {Young} E.~D.,  2014, \mn@doi [Annual Review of Earth and Planetary Sciences] {10.1146/annurev-earth-060313-054740}, \href {https://ui.adsabs.harvard.edu/abs/2014AREPS..42...45J} {42, 45}

\bibitem[\protect\citeauthoryear{{Katz}}{{Katz}}{2018}]{2018MNRAS.478L..95K}
{Katz} J.~I.,  2018, \mn@doi [\mnras] {10.1093/mnrasl/sly074}, \href {https://ui.adsabs.harvard.edu/abs/2018MNRAS.478L..95K} {478, L95}

\bibitem[\protect\citeauthoryear{{Keil}}{{Keil}}{2000}]{2000P&SS...48..887K}
{Keil} K.,  2000, \mn@doi [\planss] {10.1016/S0032-0633(00)00054-4}, \href {https://ui.adsabs.harvard.edu/abs/2000P&SS...48..887K} {48, 887}

\bibitem[\protect\citeauthoryear{{Kilic}, {Bergeron}, {Kosakowski}, {Brown}, {Ag{\"u}eros}  \& {Blouin}}{{Kilic} et~al.}{2020}]{2020ApJ...898...84K}
{Kilic} M.,  {Bergeron} P.,  {Kosakowski} A.,  {Brown} W.~R.,  {Ag{\"u}eros} M.~A.,   {Blouin} S.,  2020, \mn@doi [\apj] {10.3847/1538-4357/ab9b8d}, \href {https://ui.adsabs.harvard.edu/abs/2020ApJ...898...84K} {898, 84}

\bibitem[\protect\citeauthoryear{{Koester}}{{Koester}}{2009}]{2009A&A...498..517K}
{Koester} D.,  2009, \mn@doi [\aap] {10.1051/0004-6361/200811468}, \href {https://ui.adsabs.harvard.edu/abs/2009A&A...498..517K} {498, 517}

\bibitem[\protect\citeauthoryear{{Koester}, {G{\"a}nsicke}  \& {Farihi}}{{Koester} et~al.}{2014}]{Koester_2014}
{Koester} D.,  {G{\"a}nsicke} B.~T.,   {Farihi} J.,  2014, \mn@doi [\aap] {10.1051/0004-6361/201423691}, \href {https://ui.adsabs.harvard.edu/abs/2014A&A...566A..34K} {566, A34}

\bibitem[\protect\citeauthoryear{{Li}, {Mustill}  \& {Davies}}{{Li} et~al.}{2022}]{2022ApJ...924...61L}
{Li} D.,  {Mustill} A.~J.,   {Davies} M.~B.,  2022, \mn@doi [\apj] {10.3847/1538-4357/ac33a8}, \href {https://ui.adsabs.harvard.edu/abs/2022ApJ...924...61L} {924, 61}

\bibitem[\protect\citeauthoryear{{Lichtenberg}, {Keller}, {Katz}, {Golabek}  \& {Gerya}}{{Lichtenberg} et~al.}{2019}]{2019E&PSL.507..154L}
{Lichtenberg} T.,  {Keller} T.,  {Katz} R.~F.,  {Golabek} G.~J.,   {Gerya} T.~V.,  2019, \mn@doi [Earth and Planetary Science Letters] {10.1016/j.epsl.2018.11.034}, \href {https://ui.adsabs.harvard.edu/abs/2019E&PSL.507..154L} {507, 154}

\bibitem[\protect\citeauthoryear{{Lichtenberg}, {Dr{\k{a}}{\.z}kowska}, {Sch{\"o}nb{\"a}chler}, {Golabek}  \& {Hands}}{{Lichtenberg} et~al.}{2021}]{2021Sci...371..365L}
{Lichtenberg} T.,  {Dr{\k{a}}{\.z}kowska} J.,  {Sch{\"o}nb{\"a}chler} M.,  {Golabek} G.~J.,   {Hands} T.~O.,  2021, \mn@doi [Science] {10.1126/science.abb3091}, \href {https://ui.adsabs.harvard.edu/abs/2021Sci...371..365L} {371, 365}

\bibitem[\protect\citeauthoryear{{Lodders}}{{Lodders}}{2003}]{2003ApJ...591.1220L}
{Lodders} K.,  2003, \mn@doi [\apj] {10.1086/375492}, \href {https://ui.adsabs.harvard.edu/abs/2003ApJ...591.1220L} {591, 1220}

\bibitem[\protect\citeauthoryear{{MacLeod}, {Cantiello}  \& {Soares-Furtado}}{{MacLeod} et~al.}{2018}]{2018ApJ...853L...1M}
{MacLeod} M.,  {Cantiello} M.,   {Soares-Furtado} M.,  2018, \mn@doi [\apjl] {10.3847/2041-8213/aaa5fa}, \href {https://ui.adsabs.harvard.edu/abs/2018ApJ...853L...1M} {853, L1}

\bibitem[\protect\citeauthoryear{{Malamud} \& {Perets}}{{Malamud} \& {Perets}}{2016}]{2016ApJ...832..160M}
{Malamud} U.,  {Perets} H.~B.,  2016, \mn@doi [\apj] {10.3847/0004-637X/832/2/160}, \href {https://ui.adsabs.harvard.edu/abs/2016ApJ...832..160M} {832, 160}

\bibitem[\protect\citeauthoryear{{Malamud} \& {Perets}}{{Malamud} \& {Perets}}{2017a}]{2017ApJ...842...67M}
{Malamud} U.,  {Perets} H.~B.,  2017a, \mn@doi [\apj] {10.3847/1538-4357/aa7055}, \href {https://ui.adsabs.harvard.edu/abs/2017ApJ...842...67M} {842, 67}

\bibitem[\protect\citeauthoryear{{Malamud} \& {Perets}}{{Malamud} \& {Perets}}{2017b}]{2017ApJ...849....8M}
{Malamud} U.,  {Perets} H.~B.,  2017b, \mn@doi [\apj] {10.3847/1538-4357/aa8df5}, \href {https://ui.adsabs.harvard.edu/abs/2017ApJ...849....8M} {849, 8}

\bibitem[\protect\citeauthoryear{Malamud, Grishin  \& Brouwers}{Malamud et~al.}{2020}]{Malamud_2020}
Malamud U.,  Grishin E.,   Brouwers M.,  2020, \mn@doi [Monthly Notices of the Royal Astronomical Society] {10.1093/mnras/staa3940}

\bibitem[\protect\citeauthoryear{{Malhotra} \& {Wang}}{{Malhotra} \& {Wang}}{2017}]{2017MNRAS.465.4381M}
{Malhotra} R.,  {Wang} X.,  2017, \mn@doi [\mnras] {10.1093/mnras/stw3009}, \href {https://ui.adsabs.harvard.edu/abs/2017MNRAS.465.4381M} {465, 4381}

\bibitem[\protect\citeauthoryear{Maloney \& Gallagher}{Maloney \& Gallagher}{2010}]{Maloney_2010}
Maloney S.~A.,  Gallagher P.~T.,  2010, \mn@doi [The Astrophysical Journal] {10.1088/2041-8205/724/2/l127}, 724, L127

\bibitem[\protect\citeauthoryear{{Masiero}, {Davidsson}, {Liu}, {Moore}  \& {Tuite}}{{Masiero} et~al.}{2021}]{2021arXiv210807331M}
{Masiero} J.~R.,  {Davidsson} B. J.~R.,  {Liu} Y.,  {Moore} K.,   {Tuite} M.,  2021, \mn@doi [arXiv e-prints] {10.48550/arXiv.2108.07331}, \href {https://ui.adsabs.harvard.edu/abs/2021arXiv210807331M} {p. arXiv:2108.07331}

\bibitem[\protect\citeauthoryear{{McCoy}, {Dickinson}  \& {Lofgren}}{{McCoy} et~al.}{1999}]{1999M&PS...34..735M}
{McCoy} T.~J.,  {Dickinson} T.~L.,   {Lofgren} G.~E.,  1999, \mn@doi [Meteoritics \& Planetary Science] {10.1111/j.1945-5100.1999.tb01386.x}, \href {https://ui.adsabs.harvard.edu/abs/1999M&PS...34..735M} {34, 735}

\bibitem[\protect\citeauthoryear{{McDonald} \& {Veras}}{{McDonald} \& {Veras}}{2021}]{2021MNRAS.506.4031M}
{McDonald} C.~H.,  {Veras} D.,  2021, \mn@doi [\mnras] {10.1093/mnras/stab1906}, \href {https://ui.adsabs.harvard.edu/abs/2021MNRAS.506.4031M} {506, 4031}

\bibitem[\protect\citeauthoryear{M{\'{e} }ndez \& Rivera-Valent{\'{\i}}n}{M{\'{e} }ndez \& Rivera-Valent{\'{\i}}n}{2017}]{M_ndez_2017}
M{\'{e} }ndez A.,  Rivera-Valent{\'{\i}}n E.~G.,  2017, \mn@doi [The Astrophysical Journal] {10.3847/2041-8213/aa5f13}, 837, L1

\bibitem[\protect\citeauthoryear{{Metzger}, {Rafikov}  \& {Bochkarev}}{{Metzger} et~al.}{2012}]{2012MNRAS.423..505M}
{Metzger} B.~D.,  {Rafikov} R.~R.,   {Bochkarev} K.~V.,  2012, \mn@doi [\mnras] {10.1111/j.1365-2966.2012.20895.x}, \href {https://ui.adsabs.harvard.edu/abs/2012MNRAS.423..505M} {423, 505}

\bibitem[\protect\citeauthoryear{Miao, Li  \& Chen}{Miao et~al.}{2014}]{Miao2014}
Miao S.~Q.,  Li H.~P.,   Chen G.,  2014, \mn@doi [Journal of Thermal Analysis and Calorimetry] {10.1007/s10973-013-3427-2}, 115, 1057

\bibitem[\protect\citeauthoryear{Mustill \& Villaver}{Mustill \& Villaver}{2012}]{Mustill_2012}
Mustill A.~J.,  Villaver E.,  2012, \mn@doi [The Astrophysical Journal] {10.1088/0004-637x/761/2/121}, 761, 121

\bibitem[\protect\citeauthoryear{Narasimhan}{Narasimhan}{1999}]{https://doi.org/10.1029/1998RG900006}
Narasimhan T.~N.,  1999, \mn@doi [Reviews of Geophysics] {https://doi.org/10.1029/1998RG900006}, 37, 151

\bibitem[\protect\citeauthoryear{{O'Connor}, {Bildsten}, {Cantiello}  \& {Lai}}{{O'Connor} et~al.}{2023}]{2023ApJ...950..128O}
{O'Connor} C.~E.,  {Bildsten} L.,  {Cantiello} M.,   {Lai} D.,  2023, \mn@doi [\apj] {10.3847/1538-4357/acd2d4}, \href {https://ui.adsabs.harvard.edu/abs/2023ApJ...950..128O} {950, 128}

\bibitem[\protect\citeauthoryear{{O'Neill} \& {Palme}}{{O'Neill} \& {Palme}}{2008}]{2008RSPTA.366.4205O}
{O'Neill} H. S.~C.,  {Palme} H.,  2008, \mn@doi [Philosophical Transactions of the Royal Society of London Series A] {10.1098/rsta.2008.0111}, \href {https://ui.adsabs.harvard.edu/abs/2008RSPTA.366.4205O} {366, 4205}

\bibitem[\protect\citeauthoryear{{Pan} \& {Schlichting}}{{Pan} \& {Schlichting}}{2012}]{2012ApJ...747..113P}
{Pan} M.,  {Schlichting} H.~E.,  2012, \mn@doi [\apj] {10.1088/0004-637X/747/2/113}, \href {https://ui.adsabs.harvard.edu/abs/2012ApJ...747..113P} {747, 113}

\bibitem[\protect\citeauthoryear{{Paquette}, {Pelletier}, {Fontaine}  \& {Michaud}}{{Paquette} et~al.}{1986}]{1986ApJS...61..177P}
{Paquette} C.,  {Pelletier} C.,  {Fontaine} G.,   {Michaud} G.,  1986, \mn@doi [\apjs] {10.1086/191111}, \href {https://ui.adsabs.harvard.edu/abs/1986ApJS...61..177P} {61, 177}

\bibitem[\protect\citeauthoryear{{Paxton}}{{Paxton}}{2021}]{2021zndo...5798242P}
{Paxton} B.,  2021, {Modules for Experiments in Stellar Astrophysics (MESA)}, Zenodo, \mn@doi{10.5281/zenodo.5798242}

\bibitem[\protect\citeauthoryear{{Paxton}, {Bildsten}, {Dotter}, {Herwig}, {Lesaffre}  \& {Timmes}}{{Paxton} et~al.}{2011}]{2011ApJS..192....3P}
{Paxton} B.,  {Bildsten} L.,  {Dotter} A.,  {Herwig} F.,  {Lesaffre} P.,   {Timmes} F.,  2011, \mn@doi [\apjs] {10.1088/0067-0049/192/1/3}, \href {https://ui.adsabs.harvard.edu/abs/2011ApJS..192....3P} {192, 3}

\bibitem[\protect\citeauthoryear{{Paxton} et~al.,}{{Paxton} et~al.}{2013}]{2013ApJS..208....4P}
{Paxton} B.,  et~al., 2013, \mn@doi [\apjs] {10.1088/0067-0049/208/1/4}, \href {https://ui.adsabs.harvard.edu/abs/2013ApJS..208....4P} {208, 4}

\bibitem[\protect\citeauthoryear{{Paxton} et~al.,}{{Paxton} et~al.}{2015}]{2015ApJS..220...15P}
{Paxton} B.,  et~al., 2015, \mn@doi [\apjs] {10.1088/0067-0049/220/1/15}, \href {https://ui.adsabs.harvard.edu/abs/2015ApJS..220...15P} {220, 15}

\bibitem[\protect\citeauthoryear{{Pringle}, {Moynier}, {Savage}, {Badro}  \& {Barrat}}{{Pringle} et~al.}{2014}]{2014PNAS..11117029P}
{Pringle} E.~A.,  {Moynier} F.,  {Savage} P.~S.,  {Badro} J.,   {Barrat} J.-A.,  2014, \mn@doi [Proceedings of the National Academy of Science] {10.1073/pnas.1418889111}, \href {https://ui.adsabs.harvard.edu/abs/2014PNAS..11117029P} {111, 17029}

\bibitem[\protect\citeauthoryear{Putirka \& Xu}{Putirka \& Xu}{2021}]{article}
Putirka K.,  Xu S.,  2021, \mn@doi [Nature Communications] {10.1038/s41467-021-26403-8}, 12

\bibitem[\protect\citeauthoryear{{Rafikov}}{{Rafikov}}{2011a}]{2011MNRAS.416L..55R}
{Rafikov} R.~R.,  2011a, \mn@doi [\mnras] {10.1111/j.1745-3933.2011.01096.x}, \href {https://ui.adsabs.harvard.edu/abs/2011MNRAS.416L..55R} {416, L55}

\bibitem[\protect\citeauthoryear{{Rafikov}}{{Rafikov}}{2011b}]{2011ApJ...732L...3R}
{Rafikov} R.~R.,  2011b, \mn@doi [\apjl] {10.1088/2041-8205/732/1/L3}, \href {https://ui.adsabs.harvard.edu/abs/2011ApJ...732L...3R} {732, L3}

\bibitem[\protect\citeauthoryear{{Ricard}, {{\v{S}}r{\'a}mek}  \& {Dubuffet}}{{Ricard} et~al.}{2009}]{2009E&PSL.284..144R}
{Ricard} Y.,  {{\v{S}}r{\'a}mek} O.,   {Dubuffet} F.,  2009, \mn@doi [Earth and Planetary Science Letters] {10.1016/j.epsl.2009.04.021}, \href {https://ui.adsabs.harvard.edu/abs/2009E&PSL.284..144R} {284, 144}

\bibitem[\protect\citeauthoryear{{Schaefer} \& {Fegley}}{{Schaefer} \& {Fegley}}{2008}]{2008arXiv0801.1099S}
{Schaefer} L.,  {Fegley} Bruce J.,  2008, \mn@doi [arXiv e-prints] {10.48550/arXiv.0801.1099}, \href {https://ui.adsabs.harvard.edu/abs/2008arXiv0801.1099S} {p. arXiv:0801.1099}

\bibitem[\protect\citeauthoryear{{Scheinberg}, {Fu}, {Elkins-Tanton}  \& {Weiss}}{{Scheinberg} et~al.}{2015}]{2015aste.book..533S}
{Scheinberg} A.,  {Fu} R.~R.,  {Elkins-Tanton} L.~T.,   {Weiss} B.~P.,  2015, in , Asteroids IV.
pp 533--552, \mn@doi{10.2458/azu_uapress_9780816532131-ch028}

\bibitem[\protect\citeauthoryear{{Schlichting}, {Fuentes}  \& {Trilling}}{{Schlichting} et~al.}{2013}]{2013AJ....146...36S}
{Schlichting} H.~E.,  {Fuentes} C.~I.,   {Trilling} D.~E.,  2013, \mn@doi [\aj] {10.1088/0004-6256/146/2/36}, \href {https://ui.adsabs.harvard.edu/abs/2013AJ....146...36S} {146, 36}

\bibitem[\protect\citeauthoryear{Siebert, Sossi, Blanchard, Mahan, Badro  \& Moynier}{Siebert et~al.}{2018}]{SIEBERT2018130}
Siebert J.,  Sossi P.~A.,  Blanchard I.,  Mahan B.,  Badro J.,   Moynier F.,  2018, \mn@doi [Earth and Planetary Science Letters] {https://doi.org/10.1016/j.epsl.2017.12.042}, 485, 130

\bibitem[\protect\citeauthoryear{{Simon}, {Armitage}, {Li}  \& {Youdin}}{{Simon} et~al.}{2016}]{2016ApJ...822...55S}
{Simon} J.~B.,  {Armitage} P.~J.,  {Li} R.,   {Youdin} A.~N.,  2016, \mn@doi [\apj] {10.3847/0004-637X/822/1/55}, \href {https://ui.adsabs.harvard.edu/abs/2016ApJ...822...55S} {822, 55}

\bibitem[\protect\citeauthoryear{{Simon}, {Armitage}, {Youdin}  \& {Li}}{{Simon} et~al.}{2017}]{2017ApJ...847L..12S}
{Simon} J.~B.,  {Armitage} P.~J.,  {Youdin} A.~N.,   {Li} R.,  2017, \mn@doi [\apjl] {10.3847/2041-8213/aa8c79}, \href {https://ui.adsabs.harvard.edu/abs/2017ApJ...847L..12S} {847, L12}

\bibitem[\protect\citeauthoryear{{Staff}, {De Marco}, {Wood}, {Galaviz}  \& {Passy}}{{Staff} et~al.}{2016}]{2016MNRAS.458..832S}
{Staff} J.~E.,  {De Marco} O.,  {Wood} P.,  {Galaviz} P.,   {Passy} J.-C.,  2016, \mn@doi [\mnras] {10.1093/mnras/stw331}, \href {https://ui.adsabs.harvard.edu/abs/2016MNRAS.458..832S} {458, 832}

\bibitem[\protect\citeauthoryear{Stokes}{Stokes}{2009}]{stokes_2009}
Stokes G.~G.,  2009, Analytical Investigation.
Cambridge University Press, p. 11–75, \mn@doi{10.1017/CBO9780511702266.003}

\bibitem[\protect\citeauthoryear{{Swan}, {Farihi}, {Koester}, {Hollands}, {Parsons}, {Cauley}, {Redfield}  \& {G{\"a}nsicke}}{{Swan} et~al.}{2019}]{2019MNRAS.490..202S}
{Swan} A.,  {Farihi} J.,  {Koester} D.,  {Hollands} M.,  {Parsons} S.,  {Cauley} P.~W.,  {Redfield} S.,   {G{\"a}nsicke} B.~T.,  2019, \mn@doi [\mnras] {10.1093/mnras/stz2337}, \href {https://ui.adsabs.harvard.edu/abs/2019MNRAS.490..202S} {490, 202}

\bibitem[\protect\citeauthoryear{Trierweiler, Doyle  \& Young}{Trierweiler et~al.}{2023}]{trierweiler2023chondritic}
Trierweiler I.~L.,  Doyle A.~E.,   Young E.~D.,  2023, A Chondritic Solar Neighborhood (\mn@eprint {arXiv} {2306.03743})

\bibitem[\protect\citeauthoryear{Turner \& Wyatt}{Turner \& Wyatt}{2019}]{10.1093/mnras/stz3191}
Turner S. G.~D.,  Wyatt M.~C.,  2019, \mn@doi [Monthly Notices of the Royal Astronomical Society] {10.1093/mnras/stz3191}, 491, 4672

\bibitem[\protect\citeauthoryear{{Veras}}{{Veras}}{2020}]{2020MNRAS.493.4692V}
{Veras} D.,  2020, \mn@doi [\mnras] {10.1093/mnras/staa625}, \href {https://ui.adsabs.harvard.edu/abs/2020MNRAS.493.4692V} {493, 4692}

\bibitem[\protect\citeauthoryear{Veras \& Fuller}{Veras \& Fuller}{2019}]{10.1093/mnras/stz2339}
Veras D.,  Fuller J.,  2019, \mn@doi [Monthly Notices of the Royal Astronomical Society] {10.1093/mnras/stz2339}, 489, 2941

\bibitem[\protect\citeauthoryear{Veras, Wyatt, Mustill, Bonsor  \& Eldridge}{Veras et~al.}{2011}]{Veras_2011}
Veras D.,  Wyatt M.~C.,  Mustill A.~J.,  Bonsor A.,   Eldridge J.~J.,  2011, \mn@doi [Monthly Notices of the Royal Astronomical Society] {10.1111/j.1365-2966.2011.19393.x}, 417, 2104

\bibitem[\protect\citeauthoryear{{Veras}, {Jacobson}  \& {G{\"a}nsicke}}{{Veras} et~al.}{2014}]{2014MNRAS.445.2794V}
{Veras} D.,  {Jacobson} S.~A.,   {G{\"a}nsicke} B.~T.,  2014, \mn@doi [\mnras] {10.1093/mnras/stu1926}, \href {https://ui.adsabs.harvard.edu/abs/2014MNRAS.445.2794V} {445, 2794}

\bibitem[\protect\citeauthoryear{Veras, Leinhardt, Eggl  \& Gaensicke}{Veras et~al.}{2015}]{https://doi.org/10.48550/arxiv.1505.06204}
Veras D.,  Leinhardt Z.~M.,  Eggl S.,   Gaensicke B.~T.,  2015, Formation of planetary debris discs around white dwarfs II: Shrinking extremely eccentric collisionless rings, \mn@doi{10.48550/ARXIV.1505.06204}, \url {https://arxiv.org/abs/1505.06204}

\bibitem[\protect\citeauthoryear{{Veras}, {Higuchi}  \& {Ida}}{{Veras} et~al.}{2019}]{2019MNRAS.485..708V}
{Veras} D.,  {Higuchi} A.,   {Ida} S.,  2019, \mn@doi [\mnras] {10.1093/mnras/stz421}, \href {https://ui.adsabs.harvard.edu/abs/2019MNRAS.485..708V} {485, 708}

\bibitem[\protect\citeauthoryear{{Veras}, {McDonald}  \& {Makarov}}{{Veras} et~al.}{2020}]{2020MNRAS.492.5291V}
{Veras} D.,  {McDonald} C.~H.,   {Makarov} V.~V.,  2020, \mn@doi [\mnras] {10.1093/mnras/staa243}, \href {https://ui.adsabs.harvard.edu/abs/2020MNRAS.492.5291V} {492, 5291}

\bibitem[\protect\citeauthoryear{{Veras}, {Birader}  \& {Zaman}}{{Veras} et~al.}{2022}]{2022MNRAS.510.3379V}
{Veras} D.,  {Birader} Y.,   {Zaman} U.,  2022, \mn@doi [\mnras] {10.1093/mnras/stab3490}, \href {https://ui.adsabs.harvard.edu/abs/2022MNRAS.510.3379V} {510, 3379}

\bibitem[\protect\citeauthoryear{{Villaver} \& {Livio}}{{Villaver} \& {Livio}}{2009}]{2009ApJ...705L..81V}
{Villaver} E.,  {Livio} M.,  2009, \mn@doi [\apjl] {10.1088/0004-637X/705/1/L81}, \href {https://ui.adsabs.harvard.edu/abs/2009ApJ...705L..81V} {705, L81}

\bibitem[\protect\citeauthoryear{{Villaver}, {Livio}, {Mustill}  \& {Siess}}{{Villaver} et~al.}{2014}]{2014ApJ...794....3V}
{Villaver} E.,  {Livio} M.,  {Mustill} A.~J.,   {Siess} L.,  2014, \mn@doi [\apj] {10.1088/0004-637X/794/1/3}, \href {https://ui.adsabs.harvard.edu/abs/2014ApJ...794....3V} {794, 3}

\bibitem[\protect\citeauthoryear{{Vollstaedt}, {Mezger}  \& {Alibert}}{{Vollstaedt} et~al.}{2020}]{2020ApJ...897...82V}
{Vollstaedt} H.,  {Mezger} K.,   {Alibert} Y.,  2020, \mn@doi [\apj] {10.3847/1538-4357/ab97b4}, \href {https://ui.adsabs.harvard.edu/abs/2020ApJ...897...82V} {897, 82}

\bibitem[\protect\citeauthoryear{{Wang}, {Liu}, {Ireland}, {Brasser}, {Yong}  \& {Lineweaver}}{{Wang} et~al.}{2018}]{2018arXiv181004615W}
{Wang} H.~S.,  {Liu} F.,  {Ireland} T.~R.,  {Brasser} R.,  {Yong} D.,   {Lineweaver} C.~H.,  2018, \mn@doi [arXiv e-prints] {10.48550/arXiv.1810.04615}, \href {https://ui.adsabs.harvard.edu/abs/2018arXiv181004615W} {p. arXiv:1810.04615}

\bibitem[\protect\citeauthoryear{{Wang}, {Quanz}, {Yong}, {Liu}, {Seidler}, {Acu{\~n}a}  \& {Mojzsis}}{{Wang} et~al.}{2022}]{2022arXiv220409558W}
{Wang} H.~S.,  {Quanz} S.~P.,  {Yong} D.,  {Liu} F.,  {Seidler} F.,  {Acu{\~n}a} L.,   {Mojzsis} S.~J.,  2022, \mn@doi [arXiv e-prints] {10.48550/arXiv.2204.09558}, \href {https://ui.adsabs.harvard.edu/abs/2022arXiv220409558W} {p. arXiv:2204.09558}

\bibitem[\protect\citeauthoryear{{Wilson}, {Farihi}, {G{\"a}nsicke}  \& {Swan}}{{Wilson} et~al.}{2019}]{2019MNRAS.487..133W}
{Wilson} T.~G.,  {Farihi} J.,  {G{\"a}nsicke} B.~T.,   {Swan} A.,  2019, \mn@doi [\mnras] {10.1093/mnras/stz1050}, \href {https://ui.adsabs.harvard.edu/abs/2019MNRAS.487..133W} {487, 133}

\bibitem[\protect\citeauthoryear{{Zuckerman}, {Koester}, {Reid}  \& {H{\"u}nsch}}{{Zuckerman} et~al.}{2003}]{2003ApJ...596..477Z}
{Zuckerman} B.,  {Koester} D.,  {Reid} I.~N.,   {H{\"u}nsch} M.,  2003, \mn@doi [\apj] {10.1086/377492}, \href {https://ui.adsabs.harvard.edu/abs/2003ApJ...596..477Z} {596, 477}

\bibitem[\protect\citeauthoryear{{Zuckerman}, {Melis}, {Klein}, {Koester}  \& {Jura}}{{Zuckerman} et~al.}{2010}]{2010ApJ...722..725Z}
{Zuckerman} B.,  {Melis} C.,  {Klein} B.,  {Koester} D.,   {Jura} M.,  2010, \mn@doi [\apj] {10.1088/0004-637X/722/1/725}, \href {https://ui.adsabs.harvard.edu/abs/2010ApJ...722..725Z} {722, 725}

\makeatother
\end{thebibliography}




\appendix

\section{Adiabatic mass loss}
\label{adiabatic}

The deviation from adiabatic mass loss is quantified as:

\begin{equation}
\phi=-\frac{1}{2 \pi} \frac{\dot {M}_*}{M_{\odot}\mathrm{yr}^{-1}}\left(\frac{a}{\mathrm{AU}}\right)^{\frac{3}{2}}\left(\frac{M_*}{M_{\odot}}\right)^{-\frac{3}{2}},
\end{equation}

\noindent where $\phi \gtrsim 0.01$ indicates the break down of adiabatic assumption. 

\begin{figure}
\includegraphics[width=0.45\textwidth]{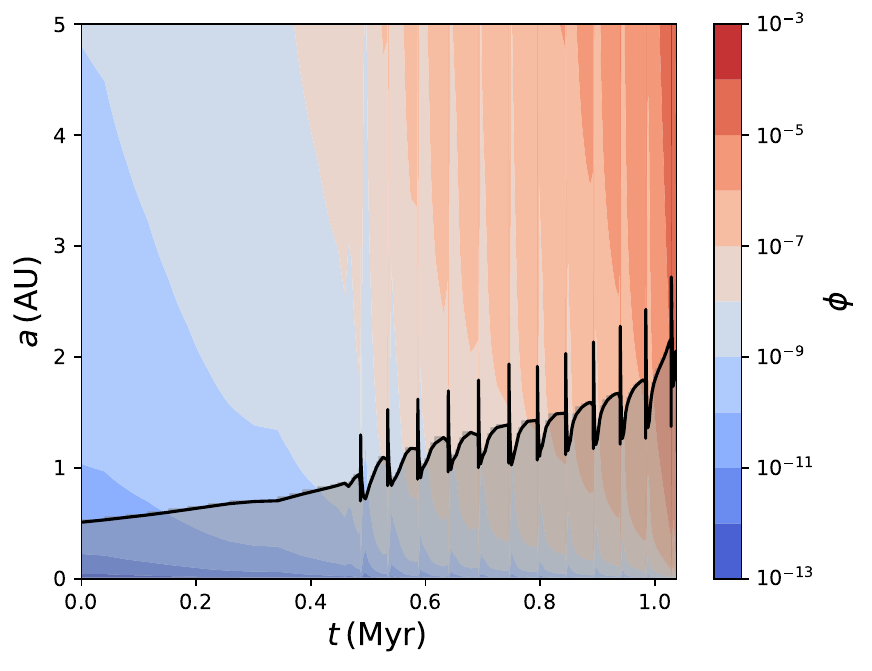}   
\caption{Parameter $\phi$ in semi-major axis-time space of a 3 solar mass star in its thermal pulsing asymptotic giant branch. The black solid curve corresponds to the radius of the host star.}
\label{adiabati plot}
\end{figure}

For close-in planetesimals within 5\,AU, even the most rapid mass loss of a 3\,$M_{\odot}$ host star (Figure \ref{adiabati plot}) is well-below the non-adiabatic threshold. The planetesimal's orbital semi-major axis must exceed $\sim 50$ (100)\,AU to enter the non-adiabatic regime during the TPAGB of the 3 (1)\,$M_{\odot}$ star.

\section{Orbital decay}

\subsection{Stellar wind drag}\label{stellar wind}

\begin{figure}
\includegraphics[width=0.45\textwidth]{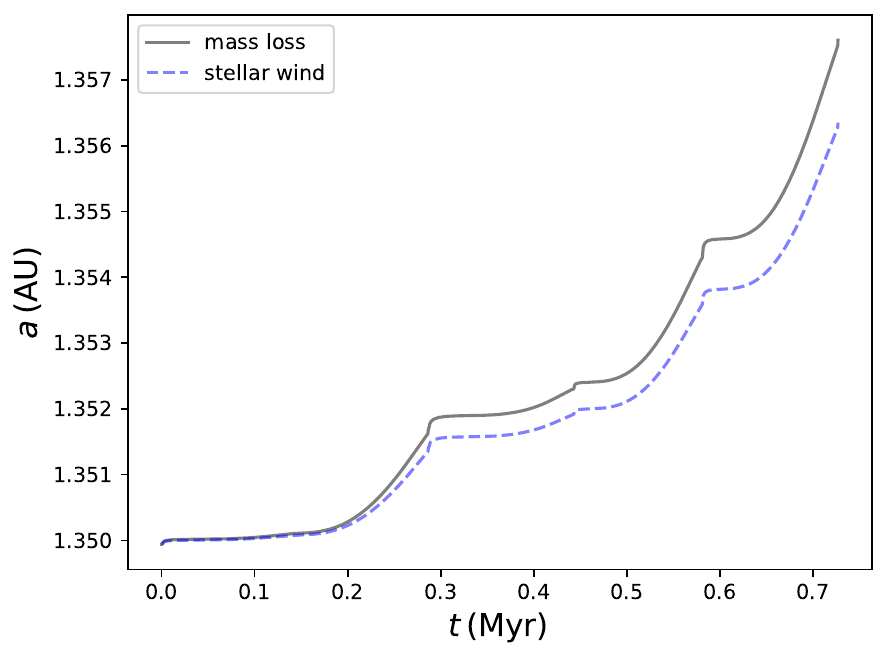}   
\caption{The orbital evolution of a planetesimal under the effect of the stellar wind drag (blue dashed line), before entering the stellar envelope since the start of TPAGB, compared to its orbital evolution dominated by stellar mass loss (black solid line). $M_*=2\,M_{\odot}$, $R_p=10$\,km, $a_0=1.3$\,AU, $e=0$, $C_d=1$.}
\label{wind drag plot}
\end{figure}

During TPAGB, mass loss becomes orders of magnitude faster, significantly increasing the density of stellar wind/circumstellar envelope. However, interactions with the circumstellar envelope can only cause an orbital decay of $\sim 0.001$\,AU ($\sim 0.1\%$ deviation), even for the smallest planetesimals ($R_p=10$\,km) close enough to enter the pulsating envelope of the 2\,$M_{\odot}$ star (Figure \ref{wind drag plot}).

\subsection{Tidal effect}\label{tide}

During the giant branches of the system, tidal interactions are dominated by stellar tides generated in the massively growing convective envelope \citep{Mustill_2012}:

\begin{equation}
\begin{aligned}
&\frac{de}{dt}|_*=-\frac{e}{36t_{\mathit{conv}}}\frac{M_{\mathit{env}}}{M_*}\left(1+\frac{M_{p}}{M_*}\right)\frac{M_{p}}{M_*}\left(\frac{R_*}{a}\right)^8\\&\times\left(\frac{5}{4}f_1-2f_2+\frac{147}{4}f_3\right),\\
&\frac{da}{dt}|_*=-\frac{a}{9t_{\mathit{conv}}}\frac{M_{\mathit{env}}}{M_*}\left(1+\frac{M_{p}}{M_*}\right)\frac{M_{p}}{M_*}\left(\frac{R_*}{a}\right)^8\\&\times\left[2f_2+e^2\left(\frac{7}{8}f_1-10f_2+\frac{441}{8}f_3\right)\right],
\end{aligned}
\end{equation}

\noindent where $M_{\mathit{env}}$ is the mass of the convective envelope and $t_{\mathit{conv}}$ is the corresponding convective time-scale:

\begin{equation}
t_{\mathit{conv}}=\left(\frac{M_{\mathit{env}}R_{\mathit{env}}^2}{\eta_F L_*}\right)^{1/3},
\end{equation}

\noindent with $R_{\mathit{env}}$ being the size of the convective envelope ($\sim R_*$), and $\eta_F\sim 3$. $f_i$ is given by:

\begin{equation}
f_i=f'min\left[1,\left(\frac{2\pi}{\sigma_ic_Ft_{\mathit{conv}}}\right)^{\gamma_F}\right],
\end{equation}

\noindent $\gamma_F\sim 2$, $f'\sim \frac{9}{2}$, $c_F\sim 1$, and $\sigma_i=i\sqrt{\frac{G(M_p+M_*)}{a^3}}$.

Meanwhile, orbital decay due to planetesimal tide is approximated by constant time lag (CTL) model. The orbital evolution of a planetesimal in 1:1 spin-orbit resonance is of the form:

\begin{equation}
\begin{aligned}
&\frac{de}{dt}|_p=\frac{81}{2Q_p'}\sqrt{\frac{G(M_p+M_*)}{a^3}}\frac{M_*}{M_p}\frac{R_p^5}{a^5}e(1-e^2)^{-\frac{13}{2}}\\&\times\left[\frac{11}{18}\frac{f_4(e)f_2(e)}{f_5(e)}-f_3(e)\right]\\&\frac{da}{dt}|_p=\frac{9}{Q_p'}\sqrt{\frac{G(M_p+M_*)}{a^3}}\frac{M_*}{M_p}\frac{R_p^5}{a^4}(1-e^2)^{-\frac{15}{2}}\\&\times\left[\frac{[f_2(e)]^2}{f_5(e)}-f_1(e)\right]
\end{aligned}
\end{equation}

\noindent where $Q_p'$ is the modified planetesimal tidal quality factor, $f$ is eccentricity functions of the form:

\begin{equation}
\begin{aligned}
&f_1(e)=1+\frac{31}{2}e^2+\frac{255}{8}e^4+\frac{185}{16}e^6+\frac{25}{64}e^8,\\
&f_2(e)=1+\frac{15}{2}e^2+\frac{45}{8}e^4+\frac{5}{16}e^6,\\
&f_3(e)=1+\frac{15}{4}e^2+\frac{15}{8}e^4+\frac{5}{64}e^6,\\
&f_4(e)=1+\frac{3}{2}e^2+\frac{1}{8}e^4\\
&f_5(e)=1+3e^2+\frac{3}{8}e^4.
\end{aligned}
\end{equation}

The evolution of semi-major axis accounting for both tidal decay and stellar mass loss can be generalised as :

\begin{equation}
\dot{a}=\dot{a}_{p,\mathit{tide}}+\dot{a}_{*,\mathit{tide}}-\frac{\dot{M}_*}{M_*}a.
\end{equation}

\begin{figure}
\includegraphics[width=0.45\textwidth]{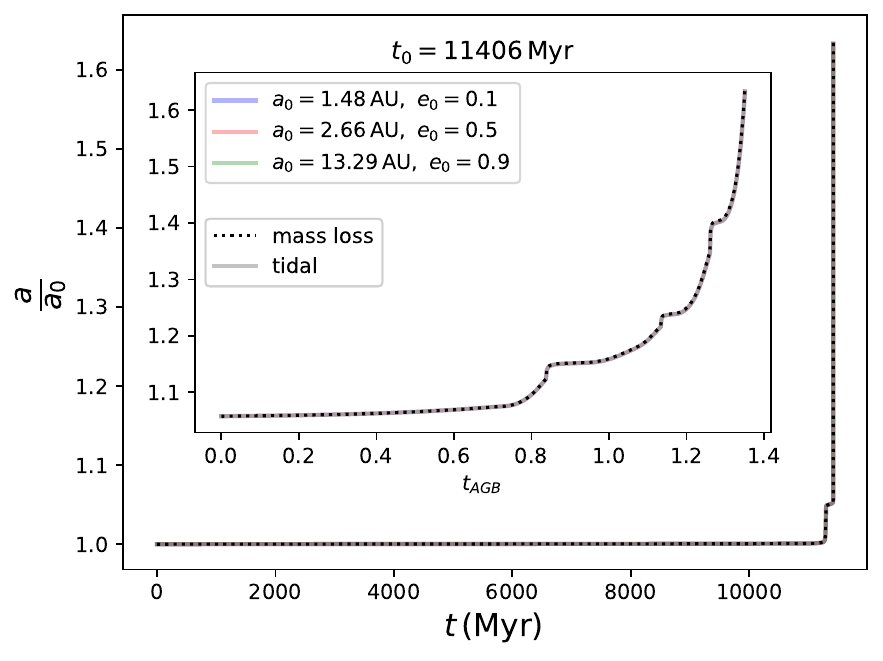}   
\caption{Comparison between the orbital evolution of a planetesimal including and excluding tidal effect from zero-age main sequence to the end of AGB life of the system. Four curves overlap with one another and cannot be visually distinguished on the plot. $R_p=300$\,km, the pericentre is fixed at 1.33\,AU (critical distance of engulfment), $M_*=1\,M_{\odot}$. The zoom-in plot corresponds to the asymptotic giant branch (AGB) with the origin of time marks the start of this phase.$t_0$ is the age of the star at the start of its AGB phase.}
\label{tidal plot}
\end{figure}

The overlap of four curves in Figure \ref{tidal plot} verifies that tidal interactions have negligible effect on planetesimal-sized bodies on eccentric orbits avoiding engulfment, and the orbital evolution of these planetesimals are dominated by stellar mass loss.

\section{Gravitational segregation}
\label{sinking}

\begin{figure}
\includegraphics[width=0.45\textwidth]{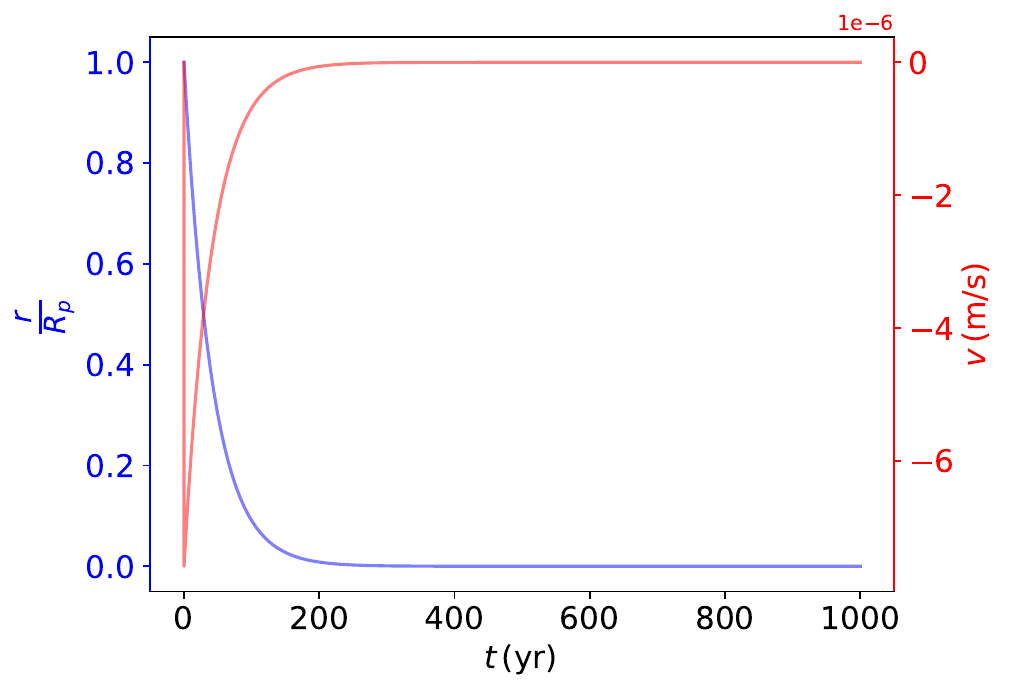}   
\caption{Radial position and velocity evolution of mm-sized fragments inside melting planetesimals. The sinking is always in the laminar regime, where Equation \ref{melt migration equation} has an analytical exponential decay solution, with the decay timescale independent of planetesimal size.}
\label{melt migration plot}
\end{figure}

The equation of motion of over-dense fragments (over-density $\Delta \rho$, radius $r_f$) is of the form \citep{stokes_2009,falkovich_2018}:

\begin{equation}\label{melt migration equation}
\ddot{r}=\begin{cases}
-\frac{\Delta\rho g}{\rho+\Delta\rho}-\frac{9 \mu \dot{r}}{2 r_f^2(\rho+\Delta \rho)} &Re\lesssim 1000 \\
-\frac{\Delta\rho g}{\rho+\Delta\rho}-\frac{3C_d \rho |\dot{r}|\dot{r}}{8 r_f(\rho+\Delta \rho)} &Re\gtrsim 1000
\end{cases} 
,
\end{equation}

\noindent where $g$ is the local gravity, $\mu$ is the dynamic viscosity, $Re=\frac{2\rho vr_f}{\mu}$ is the Reynolds number. Numerical simulations with $\mu \sim \mathrm{1\,Pa \cdot s}$ \citep{2009E&PSL.284..144R,2019E&PSL.507..154L} illustrate that flows are predominantly in the laminar regime, and mm-sized fragments sink from the surface to the core within $\sim100$\,yr (Figure \ref{melt migration plot}) $\ll$ stellar-induced melting timescale ($\sim$1\,Myr).

\section{The Yarkovsky effect and the YORP effect}
\label{Ya appendix}

The maximum effect of the Yarkovsky and the YORP effect can be approximated as \citep{2014MNRAS.445.2794V,2019MNRAS.485..708V,2022MNRAS.510.3379V,2022ApJS..262...41F}:

\begin{equation}
\label{Yar equation}
\begin{aligned}
&\left|\frac{da}{dt}\right|_{\mathit{Yar,\,max}}=\frac{3L_*}{64\pi c R_p\rho\sqrt{GM_*a}(1-e^2)},\\&\left|\frac{de}{dt}\right|_{\mathit{Yar,\,max}}=\frac{3L_*\left[e^2-2(1-e^2)\left(1-\sqrt{1-e^2}\right)\right]}{128\pi c R_p\rho\sqrt{GM_*a^3}e^3},
\end{aligned}
\end{equation}

\begin{equation}
\label{Yo equation}
\begin{aligned}
&\left|\frac{d\omega}{dt}\right|_{\mathit{YO,\,max}}=\frac{3\Phi\chi L_*}{4\pi \rho R_p^2a^2\sqrt{1-e^2}L_{\odot}},
\end{aligned}
\end{equation}

\noindent where $\Phi=10^{17}\rm\,kg\cdot m\cdot s^{-2}$ is a constant and $\chi$ quantifies the asymmetry of the planetesimal. We adopt $\chi=10^{-3}$, $M_*=2\,M_{\odot}$, $a_0=1.5$\,AU and $e=0$ and switch on the Yarkovsky effect and the YORP effect from the AGB life of the system.

\begin{figure}
\begin{center}
\includegraphics[width=0.45\textwidth]{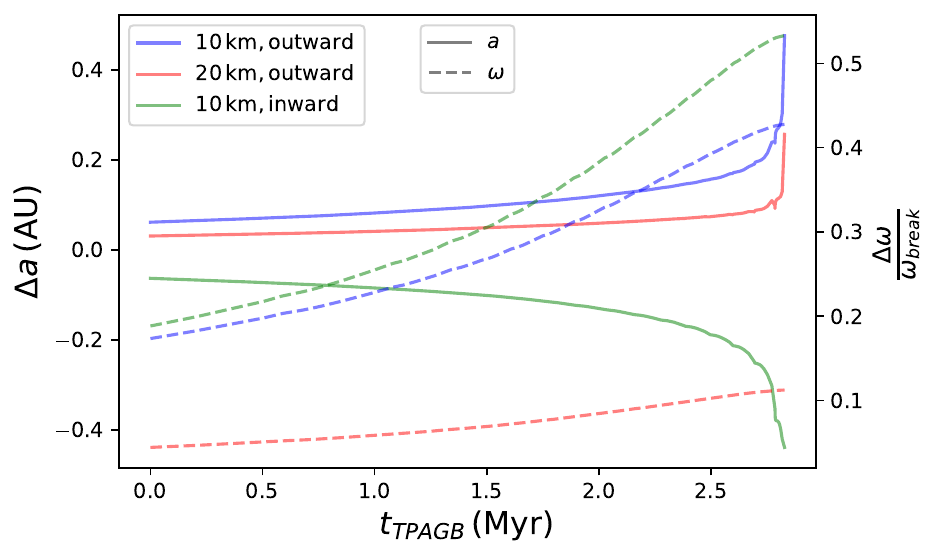}   
\caption{The maximum increase/decrease in the planetesimal's semi-major axis due to Yarkovsky effect (solid lines) and the maximum spin up of the planetesimal (dotted lines) due to YORP effect during the TPAGB phase of the host star. $\chi=10^{-3}$, $M_*=2\,M_{\odot}$, $a_0=1.5$\,AU and $e=0$. $\omega_{\mathit{break}}=\sqrt{\frac{4\pi G \rho}{3}}$. The inward drift case experiences slightly stronger Yarkovsky effect and YORP effect compared to its outward counterpart because of the stronger stellar flux $\propto \frac{L_*}{a^2}$.}
\label{Yar plot}
\end{center}
\end{figure}

As is shown in Figure \ref{Yar plot}, for the smallest planetesimals in our sample ($R_p=10$\,km ), the outward/inward Yarkovsky drift can reach $\sim 0.4$\,AU at the end of TPAGB life of the system, and the maximum YORP spin up is around $0.5\,\omega_{\mathit{break}}(=\sqrt{\frac{4\pi G\rho}{3}})$.

\bsp	
\label{lastpage}
\end{document}